\documentclass[letter,10pt]{article}
\pdfoutput=1
\usepackage{jheppub}
\usepackage[T1]{fontenc} 
\usepackage{url}
\usepackage{graphicx}
\usepackage{subcaption}
\usepackage{booktabs}
\usepackage{amsmath}
\usepackage{amssymb}
\usepackage{amstext}
\usepackage{amsfonts}
\usepackage{amsthm}
\usepackage{hyperref}
\usepackage{appendix}
\usepackage{physics}
\usepackage{aas_macros}
\usepackage[utf8]{inputenc}
\usepackage{soul}
\usepackage{hhline}
\usepackage{comment}
\usepackage{color}
\usepackage{bm}
\usepackage[normalem]{ulem}

\newcommand{\jj}{\ensuremath{\varphi}}
\newcommand{\fv}{\ensuremath{{\rm fv}}}
\newcommand{\tv}{\ensuremath{{\rm tv}}}
\newcommand{\crit}{\ensuremath{{\rm crit}}}

\newcommand{\com}{\ensuremath{{\rm \scriptscriptstyle COM}}}

\newcommand{\osc}{\ensuremath{{\rm osc}}}
\newcommand{\N}{\ensuremath{{\rm N}}}
\newcommand{\eff}{\ensuremath{{\rm eff}}}

\def\eqref#1{(\ref{#1})}

\usepackage{tabularx}
\newcolumntype{C}{>{\centering\arraybackslash}X}
\newcolumntype{R}{>{\raggedleft\arraybackslash}X}

\newcommand{\ie}{i.e.\ }
\newcommand{\eg}{e.g.\ }

\newcommand{\perimeter}{Perimeter Institute for Theoretical Physics,\\31 Caroline St N, Waterloo, ON N2L 2Y5, Canada}

\newcommand{\UW}{Department of Physics and Astronomy, University of Waterloo,\\Waterloo, ON N2L 3G1, Canada}

\newcommand{\york}{Department of Physics and Astronomy, York University,\\Toronto, ON M3J 1P3, Canada}

\newcommand{\mcmaster}{Department of Physics and Astronomy, McMaster University,\\Hamilton, Ontario, L8S 4M1, Canada}

\def\eqref#1{(\ref{#1})}

\title{\LARGE Bubble velocities and oscillon precursors in first-order phase transitions}

\author[a,b]{Dalila~P\^irvu,}

\author[a,c]{Matthew~C.~Johnson}

\author[a,d]{and Sergey~Sibiryakov}
 
\affiliation[a]{\perimeter}
\affiliation[b]{\UW}
\affiliation[c]{\york}
\affiliation[d]{\mcmaster}

\emailAdd{dpirvu@perimeterinstitute.ca}
\emailAdd{mjohnson@perimeterinstitute.ca}
\emailAdd{ssibiryakov@perimeterinstitute.ca}

\abstract{Metastable `false' vacuum states are an important feature of the Standard Model of particle physics and many theories beyond it. Describing the dynamics of a phase transition out of a false vacuum via the nucleation of bubbles is essential for understanding the cosmology of vacuum decay and the full spectrum of observables. In this paper, we study vacuum decay  by numerically evolving ensembles of field theories in 1+1 dimensions from a metastable state. We demonstrate that for an initial Bose-Einstein distribution of fluctuations, bubbles form with a Gaussian spread of center-of-mass velocities and that bubble nucleation events are preceded by an oscillon -- a long-lived, time-dependent, pseudo-stable configuration of the field. Defining an effective temperature from the long-wavelength amplitude of fluctuations in the ensemble of simulations, we find good agreement between theoretical finite temperature predictions and empirical measurements of the decay rate, velocity distribution and critical bubble solution. We comment on the generalization of our results and the implications for cosmological observables.}

\begin{document}
\maketitle
\flushbottom

\section{Introduction}\label{sec:Intro}

Quantum field theories with a metastable `false' vacuum can undergo vacuum decay -- a first-order phase transition to a lower-energy `true' vacuum via the formation, expansion, and coalescence of bubbles. Many extensions of the Standard Model (SM) of particle physics feature metastable vacuum states, and predict that vacuum decay occurred in the early Universe. The current SM vacuum itself could be metastable, with vacuum decay in our future. The full dynamical picture of vacuum decay is crucial to fully understand the physical implications of these theories and test them with existing and future cosmological observations.

Vacuum decay can occur at zero temperature via quantum tunneling ~\cite{Coleman:1977py, Callan:1977pt}, at finite temperature due to thermal fluctuations~\cite{Linde:1980tt, Linde:1981zj}, or more generally from an excited state above the false vacuum~\cite{PhysRevD.20.3168}. Euclidean instanton techniques are typically used to predict the rate per unit time per unit volume of bubble formation as well as the properties of the critical bubble -- the most probable expanding bubble to form from the quantum or thermal ensemble. However, Euclidean techniques cannot be used to describe the dynamics of the phase transition, which is ultimately required to connect with observables. One approach to a dynamical description of vacuum decay is to numerically evolve ensembles of field theory simulations, and directly measure ensemble-averaged observables of interest. This procedure, which we adopt here, is referred to as the Classical-Statistical approximation~\cite{Berges_2007, Aarts_1998, Millington:2020vkg}, stochastic approach~\cite{Linde_1992, Felder_2008, Frolov_2008, Khlebnikov:1996mc, Khlebnikov:1996wr, Khlebnikov:1996zt}, or the truncated Wigner approximation~\cite{CWGardiner_2002} depending on the context and field of study. Early work in this direction was performed for topological solitons and vacuum transitions at finite temperature in~\cite{Grigoriev:1988bd, Grigoriev:1989je, Grigoriev:1989ub, Ambjorn:1990wn, PhysRevD.47.R2168}. More recently, such an approach was used to study vacuum decay at zero temperature in the semiclassical limit~\cite{braden-newsemiclassical}, and further explored in~\cite{Blanco-Pillado_2019, braden-massreno, Hertzberg:2019wgx, Hertzberg:2020tqa, Tranberg:2022noe, Wang:2019hjx, batini2023realtime}. Other real-time perspectives on vacuum decay include~\cite{Turok:2013dfa, Cherman:2014sba, Bramberger:2016yog, Blum:2023wnb, Nishimura:2023dky, Hayashi:2021kro, Shkerin:2021zbf}.

\begin{figure}[!]
    \centering
    \includegraphics[width=0.6\textwidth]{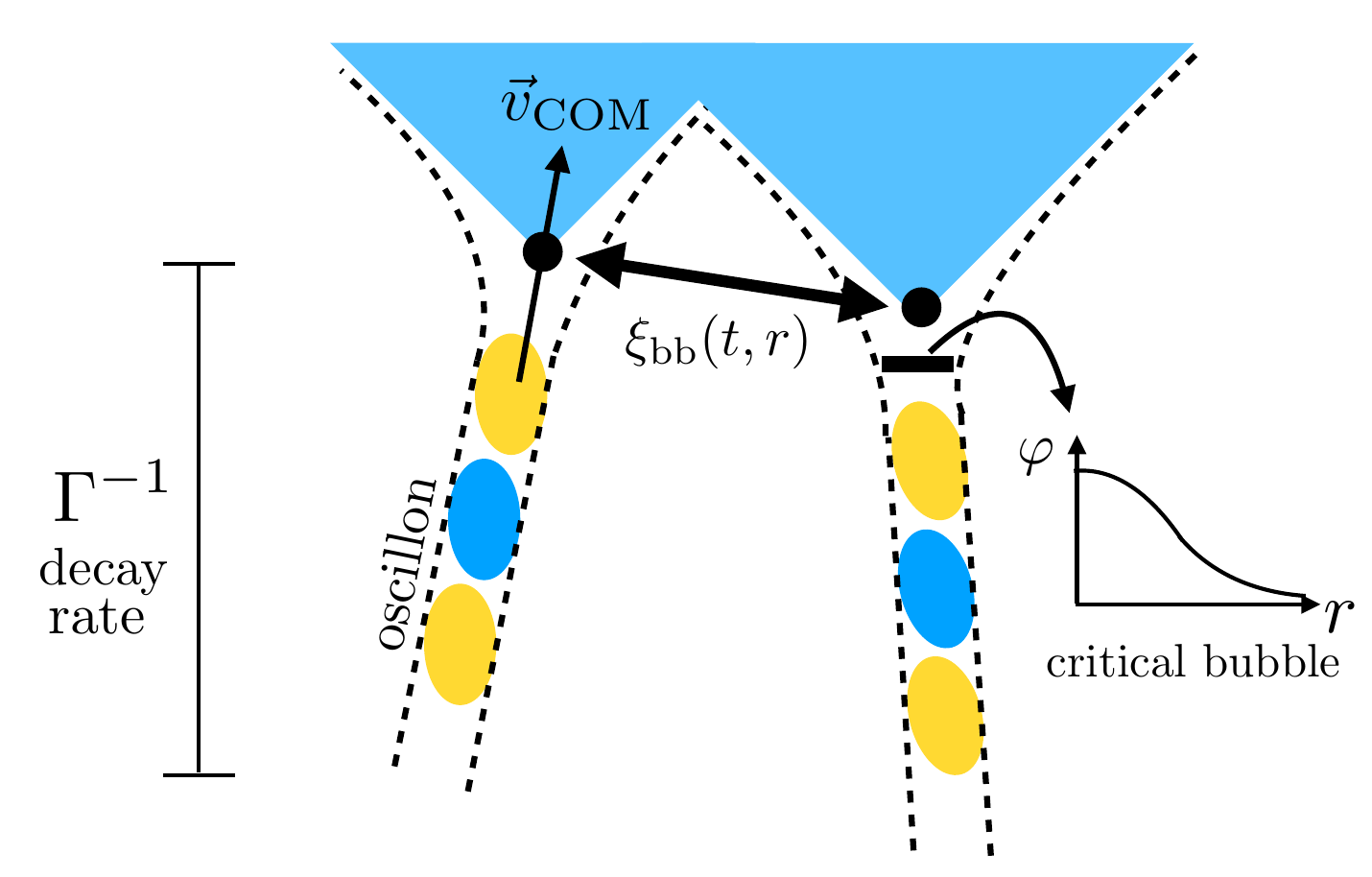}
\caption{Observables in vacuum decay. The basic observables in vacuum decay are the decay rate $\Gamma$ and the critical bubble profile $\jj(r)$, which can both be predicted from the instanton techniques 
and measured from bubbles formed in real-time simulations. Observables beyond the decay rate include the bubble-bubble correlation function $\xi_{bb} (t,r)$ that was explored in~\cite{Pirvu:2021roq}, as well as the center-of-mass velocity $\vec{v}_\com$ of nucleated bubbles and oscillon precursors that we investigate in this work.}
    \label{fig:bubble_cartoon}
\end{figure}

In this paper, we employ a classical stochastic description of vacuum decay of a single real scalar field with an initial Bose-Einstein distribution of fluctuations in 1+1 dimensions. Our focus is on identifying new observable phenomena that can only be accessed through a real-time approach. One such example is the clustering of bubble nucleation sites -- the consequence of a non-trivial bubble-bubble correlation function~\cite{Pirvu:2021roq, DeLuca:2021mlh}. This is a quantity that cannot be easily predicted using the Euclidean instanton formalism. Here, we explore additional observable phenomena by developing a set of algorithms to analyze in detail the properties of bubbles, before, during, and after nucleation. Our qualitative results are summarized in Fig.~\ref{fig:bubble_cartoon}. Defining an empirical temperature for infrared modes on the lattice, the observed decay rate is consistent with the instanton prediction at this effective temperature. This is somewhat surprising as the input Bose-Einstein distribution is not the true thermal equilibrium state of the field, and thermalization is an extremely slow process in 1+1 dimensions. Nevertheless, we find throughout that the predictions of a thermal ensemble describe our empirical measurements well.

Studying individual nucleation events, we find that bubbles never form at rest. We measure the distribution of the bubble center-of-mass velocity, which has a variance determined by the effective temperature for infrared modes and the energy of the critical bubble. An accurate measurement of the center-of-mass velocity allows us to stack nucleation events in their rest frame to determine the critical bubble and its time evolution. The measured critical bubble is consistent with the thermal Euclidean instanton prediction. Focusing on the field configuration prior to nucleation, we confirm the prediction of Ref.~\cite{Aguirre:2011ac} that the most likely formation channel for bubbles includes an oscillon precursor. Oscillons are long-lived time-dependent field configurations~\cite{MAKHANKOV1981344} arising in scalar field theories with anharmonic potentials (see \eg~\cite{Amin:2010jq, Johnson:2008se}), whose role in vacuum decay has been discussed previously in Refs.~\cite{Gleiser:1991rf, Gleiser:1993pt, PhysRevLett.94.151601, Gleiser:2007ts,Aguirre:2011ac}. The existence of bubble precursors and a center-of-mass velocity distribution could only have been confirmed with real-time description of vacuum decay, and open the door to further investigations using similar techniques. Further, we speculate that these features of vacuum decay can have observable implications for early universe phenomenology.

There have been a variety of recent efforts to perform experimental simulations of false vacuum decay at low temperature using cold atom systems~\cite{Billam:2018pvp, Billam:2020xna, Billam:2021nbc, Billam:2021psh, Billam:2022ykl, jenkins2023tabletop, Braden:2017add, Braden:2019vsw, Fialko:2014xba, Fialko:2016ggg, Opanchuk:2013lgn, PRXQuantum.2.010350, Zenesini:2023afv}, quantum annealers~\cite{PRXQuantum.2.010349}, spin chains~\cite{Lagnese_2021, lagnese2023detecting}, and (for the related Schwinger process) quantum computers~\cite{Martinez_2016, Xu:2021tey}. The detailed properties of vacuum decay described above will be important observables for these experiments. In particular, cold atom simulations of vacuum decay have recently been performed~\cite{Zenesini:2023afv}, with further results expected in the near future~\cite{jenkins2023tabletop}. Through comparing these observations with simulation and other real-time theoretical descriptions, we hope to learn a great deal about the fundamentals of vacuum decay.

The paper is organized as follows. In Section~\ref{sec:InstantonSection} we give an overview of the Euclidean instanton description of thermal vacuum decay and the properties of the critical bubble solution. In Section~\ref{sec:CodeDescription} we introduce our numerical tools. Section~\ref{sec:ResultsObservables} introduces several new observables in false vacuum decay and describes in detail the computational methods that were used to extract them. We discuss the role of field fluctuations for thermalization in Section~\ref{subsec:powespec_evol} and measure the effective temperature and mass of the field about the false vacuum. We measure the decay rate from lattice simulations and compare it to the predictions of the thermal Euclidean theory in Section~\ref{subsec:decayrate}. In Section~\ref{subsec:COMvelocitydistrib} we measure the distribution of the center-of-mass velocities of nucleated bubbles in ensembles of simulations. In 
Section~\ref{subsec:averagebubble} we stack many nucleation events to determine the ensemble-averaged most likely bubble configuration directly from the simulations. In Section~\ref{subsec:Precursors} we show that bubble nucleation events are preceded by oscillons. In Section~\ref{subsec:critical_energy} we verify the consistency of several measurements of the critical bubble energy.
Finally, we comment on the implication of these new observables for early universe scenarios in Section~\ref{sec:Conclusion}. We assume the units with $c = 1$ and work in the limit $\hbar\to 0$, unless stated otherwise.

\section{Euclidean computation of the decay rate at finite temperature}\label{sec:InstantonSection}

\begin{figure}[!]
    \centering
    \includegraphics[width=0.4\textwidth]{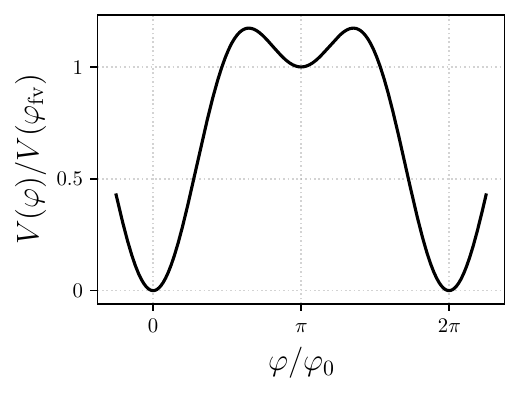}
    \caption{Potential for the relativistic field $\jj$ given by Eq.~\eqref{eq:potential} where $\lambda=1.5$. The field starts off in the false vacuum centered at $\jj_\fv=\pi \jj_0$ and can decay to either $\jj_\tv=0$ or $\jj_\tv=2\pi \jj_0$.}
    \label{fig:potential}
\end{figure}

Consider a scalar field theory in $1+1$-dimensions:
\begin{equation}\label{eq:lagrangian}
\begin{aligned}
    \mathcal{L} &= \frac{1}{2} \dot{\jj}^2-\frac{1}{2}\left(\partial_r \jj\right)^2-V(\jj).
\end{aligned}
\end{equation}
We will be focused on potentials $V(\jj)$ with a high-energy `false' minimum separated by a barrier from a low-energy `true' minimum. The form of the potential is otherwise arbitrary, although for specificity in this paper we focus on a potential of the form:
\begin{equation}\label{eq:potential}
V(\jj) = V_0 \left[ -\cos \left( \frac{\jj}{\jj_0} \right) + \frac{\lambda^2}{2} \sin^2\left( \frac{\jj}{\jj_0} \right) \right].
\end{equation}
When the parameter $\lambda>1$, the potential $V(\jj)$ has an infinite sequence of local minima at $\jj_\fv = (2n+1)\pi \jj_0$ alternating with global minima at $\jj_\tv = 2n\pi \jj_0, n\in \mathbb{Z}$. Fig.~\ref{fig:potential} illustrates the potential for the case where $\lambda=1.5$. If the field begins everywhere in the false vacuum minimum, then either through quantum mechanical or thermal effects, bubbles containing the true vacuum phase will form, expand, and coalesce -- this is false vacuum decay~\cite{Callan:1977pt, Coleman:1977py, Linde:1980tt, Linde:1981zj}.

In thermal equilibrium at some temperature $T$, the decay rate can be computed using Euclidean instanton techniques~\cite{Linde:1980tt, Linde:1981zj}. The goal is to find saddle points of the Euclidean action (equivalently the partition function) corresponding to critical bubble solutions that interpolate between a false vacuum initial condition and a true vacuum final condition -- a bubble. Beyond the temperature, the most important scale in the problem is the characteristic (Euclidean) size of the bubbles. This can generically be estimated as $r \sim 1/\sqrt{\partial_\jj^2 V(\jj_{\fv})} \equiv m^{-1}$ (it could be far larger in the thin-wall limit, but we do not consider such cases here). The statistical mechanics of the field at finite temperature can be described by a field theory in Euclidean space with a time variable that has period $\beta \equiv \hbar/T$. For $T \ll \hbar m$, the bubble is far smaller than the size of the time dimension, and the full (in our case 2 dimensional) Euclidean solution must be used -- this is the solution appropriate to describe quantum mechanical formation of bubbles~\cite{Callan:1977pt, Coleman:1977py}. In the opposite limit where $T \gg \hbar m$, the bubbles are far larger than the size of the time dimension, and the solutions are independent of Euclidean time. 

The relevant saddle point solution $\jj_{\crit} (r)$ (also known as the `critical bubble') in the high-temperature limit is found by solving
\begin{equation}
\label{eq:statsphaleron}
    \partial_r^2 \jj_{\crit} = \partial_\jj V(\jj_{\crit}),
\end{equation}
with the boundary conditions $\jj(r \rightarrow \infty) = \jj_{\fv}$ and $\partial_r \jj ({r \to 0}) = 0$. We disregard thermal corrections to the potential which are small for the case of a single scalar field studied in this work. Note, however, that they may be important if the decaying field couples to other species. The prediction for the decay rate per unit time per unit volume in our 1+1 dimensional system is
\begin{equation}\label{eq:decay_rate_bounce}
    \Gamma = A B^{1/2} e^{-B},
\end{equation}
where the critical bubble action (divided by $\hbar$) is given by
\begin{equation}
\label{eq:Bbubble}
    B = \frac{1}{T} \int \dd r \left[ \frac{1}{2} \left( \frac{\partial \jj_{\crit}}{\partial r}\right)^2 + V(\jj_{\crit}) \right] = \frac{E_\crit}{T}.
\end{equation}
In expression Eq.~\eqref{eq:decay_rate_bounce} the exponential term is the solution to the saddle point approximation to the path integral with one periodic dimension of size $\hbar/T$, while the factor $A$ encompasses the effects of fluctuations around this solution. The factor of $B^{1/2}$ comes from integrating out the shift-symmetric degree of freedom. See \eg~\cite{weinberg_2012, reviewthermaldrate} for a review of thermal instanton theory.

\begin{figure}[t!]
    \centering
    \includegraphics[width=0.8\textwidth]{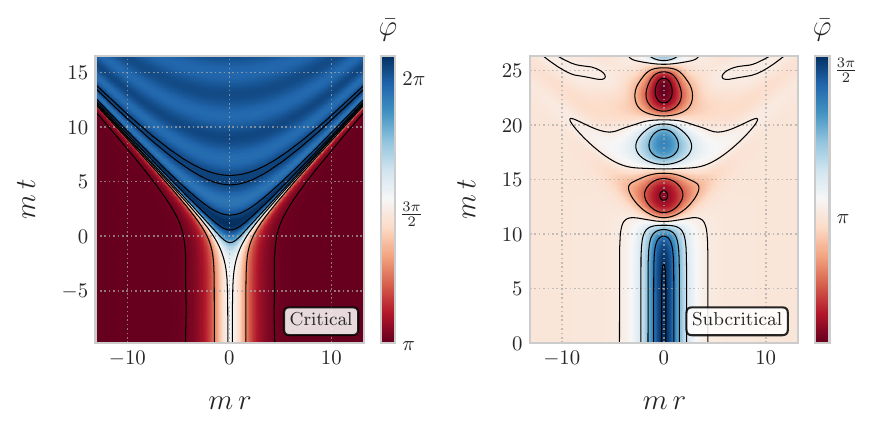}
    \caption{The time evolution of the critical bubble solution to Eq.~\eqref{eq:instEOM}. The critical bubble is in unstable equilibrium between expansion and collapse, and small changes to the initial condition determine the outcome. Here we show both outcomes, where the black contours denote constant field amplitude. Lengths and times are measured in terms of $m \equiv \sqrt{\partial_\jj^2 V(\jj_{\fv})}$; the normalized field is $\bar{\jj} \equiv \jj/\jj_0$. The false vacuum is located at $\bar{\jj} = \pi$; the true vacuum is located at $\bar{\jj} = 2\pi$. \emph{Left:} After loitering near the critical solution, the field evolves to the true vacuum inside an expanding bubble. The constant-field surfaces composing the expanding wall and the bubble interior asymptote to timelike and spacelike hyperboloids, respectively. \emph{Right:} The field configuration is slightly subcritical (the central value of the field is altered by one part in $10^4$ as compared with the left panel), so after a brief loitering period, it collapses into an oscillon -- a long-lived compact oscillating field configuration.}
    \label{fig:bare_sphaleron_and_subcritical}
\end{figure}

The critical bubble $\jj_{\crit}$ is also a static solution to the Lorentzian equation of motion
\begin{equation}\label{eq:instEOM}
    -\partial_t^2 \jj_{\crit} + \partial_r^2 \jj_{\crit} = \partial_\jj V(\jj_{\crit}).
\end{equation}
 It is unstable to either growth or collapse under small perturbations. For illustration, in Fig.~\ref{fig:bare_sphaleron_and_subcritical} we show the time evolution of a slightly rescaled configuration $\jj_\crit(r, t=0)$ under the equations of motion Eq.~\eqref{eq:instEOM}. The field profile at $t=0$ for the image on the left is the solution where the central value of the field in the bubble is $\jj(r=0) = 4.8238 \jj_0$. The stationary field profile remains nearly static for a finite amount of time, and then expands into the true vacuum. Once the bubble wall begins to expand, the surfaces of constant field follow timelike hyperboloids $\sqrt{r^2-(t-t_0)^2} = {\rm const}$, while the surfaces of constant field inside the bubble follow spacelike hyperboloids $\sqrt{(t-t_0)^2-r^2} = {\rm const}$. The spontaneous generation of hyperbolic symmetry in expanding thermal bubbles was studied in detail in Ref.~\cite{Aguirre:2008wy}, where it was shown to be a generic phenomenon. On the right, we change the central value of the field by one part in $10^4$ choosing $\jj(r=0) = 4.8237\jj_0$. This yields a solution that collapses, forming an oscillon. The collapse of subcritical bubbles into oscillons was first discussed in Ref.~\cite{Gleiser:1993pt}.

\section{Lattice simulations}\label{sec:CodeDescription}

We employ a stochastic/classical-statistical numerical approach to study the real-time dynamics of vacuum decay. We approximate the ensemble-averaged field dynamics by the non-linear classical time evolution of many realizations of initial conditions for the field drawn from a distribution over phase space centered on the false vacuum. In the simulation of each realization, bubbles of true vacuum form due to the field dynamics -- each nucleation event is different in detail. The formation history and evolution of bubbles can be recorded to obtain an ensemble whose statistical properties can be investigated in detail. For a classical distribution over phase space, this prescription is exact, at least to the extent that a sufficient number of realizations are drawn to properly sample the possible dynamics. It has been suggested that this prescription can also capture quantum effects to leading order in $\hbar$~\cite{braden-newsemiclassical}. For additional recent discussions of the stochastic/classical-statistical approach in the semiclassical limit see Refs.~\cite{braden-newsemiclassical, braden-massreno, Millington:2020vkg, batini2023realtime, Hertzberg:2019wgx, Epelbaum:2014yja}.

In this paper, we focus on a single scalar field in 1+1 dimensions. We consider initial conditions where the `occupation number' in each Fourier mode is given by the Bose-Einstein distribution:\footnote{This is a classical analog of the occupation number. It is related to the true occupation number $\hat n_k$ by $n_k=\hbar \hat n_k$. So, $\hat n_k$ diverges in the classical limit $\hbar\to 0$ with $n_k$ fixed.}
\begin{equation}\label{eq:bosedist}
    n_k = \frac{1}{e^{\omega_k/T} - 1}, \quad \omega_k^2 = m^2 + k^2.
\end{equation}
Using the potential Eq.~\eqref{eq:potential}, we set the scalar field mass $m^2$ equal to the curvature around the false vacuum $m^2 = V^{\prime \prime}\left(\bar{\jj}_\fv\right)$. The field and conjugate momentum have the spectra:
\begin{equation}\label{eq:power_spectra}
    \ev{ \delta \jj_k^* \delta \jj_{k^{\prime}} } = \frac{n_k}{\omega_k} \; \delta\left(k-k^{\prime}\right), \quad
    \ev{ \delta \Pi_k^* \delta \Pi_{k^{\prime}} } = n_k \omega_k \; \delta\left(k-k^{\prime}\right), \quad \ev{ \delta \jj_k \delta \Pi_{k^{\prime}}^* } = 0.
\end{equation}
Throughout this work we will use the $\ev{\cdot}$ notation to denote an ensemble average unless otherwise stated. Initial configurations of the field and conjugate momentum are random draws from a multivariate Gaussian for each mode with covariance Eq.~\eqref{eq:power_spectra}. 

Let us make an important comment. The choice of the Bose-Einstein distribution for the initial conditions is convenient since it ensures approximate equipartition of energy among long modes relevant for bubble nucleation, while exponentially cutting off the power in the ultraviolet, thereby reducing the sensitivity to the lattice spacing. However, it {\em does not} represent an equilibrium thermal distribution of the classical field theory which we simulate. In a free theory, the equilibrium would correspond to the Rayleigh-Jeans spectrum which has a significant power in the ultraviolet and is numerically challenging.
Presence of the field self-interaction further complicates the definition of the equilibrium state -- in fact, makes it strictly speaking impossible for the dynamics around a {\em metastable} false vacuum.\footnote{The Rayleigh-Jeans spectrum was adopted in the first numerical studies of non-perturbative processes in classical field theory in $1+1$ dimensions~\cite{Grigoriev:1988bd, Grigoriev:1989je, Grigoriev:1989ub}. Other previous works~\cite{PhysRevD.47.R2168, PhysRevD.48.2838, STRUMIA1999482, GLEISER199869} either explicitly introduced a heat bath, a thermalization phase preceding a sudden change in the potential (a quench), or incorporated the effect of a heat bath into effective equations of motion. This introduces additional complexity and model assumptions that are avoided in our approach -- we simply have an ensemble of closed systems with an initial spectrum of fluctuations.}

On the other hand, thermalization is extremely slow in $1+1$ dimensions~\cite{Boyanovsky-thermalization}. This provides an opportunity of studying in real time false vacuum decay from non-equilibrium states -- a process inaccessible with Euclidean methods (see~\cite{Shkerin:2021zbf} for the generalization of the instanton techniques for this case). The initial state with the spectrum~\eqref{eq:bosedist} should be viewed as an example of such non-equilibrium configurations. Remarkably, we will see that nucleation of the true vacuum bubbles in this state still admits an approximate thermal description, albeit with an effective temperature $T_\eff$ different from the parameter $T$ in the initial Bose-Einstein distribution~\eqref{eq:bosedist}. We give more details on the spectrum and its evolution below. We stress that our focus in this paper is not on defining a precise physically motivated set of initial conditions, but rather the identification of new observables in vacuum decay beyond the decay rate. The quantitative predictions for observables in specific early universe or experimental scenarios can be obtained by extending our results to different choices for the initial state.

In our simulations we use the numerical code described in Refs.~\cite{braden-massreno, braden-newsemiclassical, Pirvu:2021roq}. Time evolution is performed using a 10th order Gauss-Legendre integrator~\cite{Butcher:1964, Braden:2014cra}. Spatial derivatives are computed using pseudo-spectral methods. The spatial dimension is periodic. To ensure the accuracy of our simulations, we performed a number of convergence tests verifying that the spacetime coordinates of the decay events remain the same for a range of lattice spacings, $\dd r$, and integration time steps, $\dd t$. The details of these tests are given in Appendix~\ref{app:A}.

The mean field and conjugate momentum at initialization are chosen precisely at the false vacuum, while fluctuations are sampled stochastically in each realization around these values according to $\bar{\jj}(r, t=0) = \pi + \delta \bar{\jj}(r)$ and $\bar{\Pi}(r, t=0) = \delta \bar{\Pi}(r)$, where
\begin{equation}\label{eq:initialspectra}
\begin{gathered}
    \delta \bar{\jj}(r) = \frac{1}{\jj_0}\frac{1}{\sqrt{L}} \sum_{k} \left[\frac{\hat{\alpha}_k}{\sqrt{\omega_k}} \frac{e^{i k r}}{\sqrt{e^{\omega_k/T}-1}} +\text { c.c. }\right], \\
    \delta \bar{\Pi}(r) = \frac{1}{\jj_0}\frac{1}{\sqrt{L}} \sum_{k} \left[\hat{\beta}_k \sqrt{\omega_k} \frac{e^{i k r}}{\sqrt{e^{\omega_k/T}-1}} +\text { c.c. }\right].
\end{gathered}
\end{equation}
The complex random deviates $\hat{\alpha}_k$ and $\hat{\beta}_k$ are drawn from a Gaussian distribution of unit variance. Each realization of the initial conditions is then evolved using equations of motion
\begin{equation}\label{eq:code_eom}
    \begin{gathered}
        \frac{\mathrm{d} \bar{\jj}}{\mathrm{d} t}=\bar{\Pi}, \\
        \frac{\mathrm{d} \bar{\Pi}}{\mathrm{d} t}=\nabla^2 \bar{\jj}-\frac{V_0}{\jj_0^2} \left[ \sin (\bar{\jj})+\frac{\lambda^2}{2} \sin (2 \bar{\jj}) \right].
    \end{gathered}
\end{equation}
The parameter $\jj_0$ controls the width of the potential, while $V_0$ controls its height. The potential and lattice parameters are fixed throughout this work. The only parameter we vary across ensembles is $T$ in the initial spectrum~\eqref{eq:bosedist} which can take one of the following values: $T = \{0.9 m, m, 1.1 m, 1.2 m \}$. The potential is defined by $\lambda=1.5$ and $V_0/\jj_0^2 = 0.008$, with $\jj_0=2\pi/4.5$. The physical size of the lattice is $L = 50\sqrt{2} \jj_0 / \sqrt{V_0}$ and we sample $N=1024$ points. This gives a lattice unit $\dd r = L/N \approx 0.77$ and an IR scale $\dd k = 2\pi/L \approx 7.95\times10^{-3}$. Wave-numbers $k$ run from $k_{\rm IR} = \dd k$ to $k_{\rm UV} = \dd k N/2$. We perform a total of $4000$ simulations at each value of $T$, and monitor the time evolution up to at most $5 L$, \ie five lattice crossing times, or until the field has completed the phase transition into the true vacuum. The discrete time step for the integration procedure is $\dd t = \dd r / 16$. The speed of light is fixed in simulations to $c = \dd r/\dd t_{\rm out} = 1$, where $\dd t_{\rm out}=16\dd t$ is the interval over which the data are output to produce spacetime diagrams such as \eg Fig.~\ref{fig:bare_sphaleron_and_subcritical}. We typically express length and time scales as a function of the mass which is fixed to $m(\jj) = V_0 \jj_0^{-2} \left( \lambda^2 - 1 \right) = 0.1$ in the dimensionless code units. The critical bubble energy is $E_\crit \approx 0.33$ in the code units, or $E_\crit \approx 3.3m$. We also use the notation $\bar{\jj}=\jj/\jj_0$, where $\bar{\jj}$ is the normalized field amplitude. A summary of the most relevant parameters is given in Table~\ref{table:1}. In the rest of this work, all dimensionful quantities are plotted in terms of the bare mass scale $m$.

For future reference, we denote the expectation values for the variance of field and momentum fluctuations by $\sigma_\jj^2 \equiv \ev{ \delta \jj^2 }$ and $\sigma_\Pi^2 \equiv \ev{ \delta \Pi^2 }$, where the average is taken over many different realizations. At the initial moment of time we can write them in terms of the Bose-Einstein distribution~\eqref{eq:bosedist} as
\begin{equation}\label{eq:vairances_with_distrib}
    \jj_0^2 \sigma_\jj^2 = \frac{1}{L} \sum_k \frac{n_k}{\omega_k}, \quad \jj_0^2 \sigma_\Pi^2 = \frac{1}{L} \sum_k n_k \omega_k.
\end{equation}
From low to high values of $T$, the ratio $\sigma_{\jj} = \sigma_{\bar{\jj}}/\jj_0$ gives $\approx \{ 0.20, 0.22, 0.24, 0.26\}$, whereas $\sigma_{\Pi} = \sigma_{\bar{\Pi}}/\jj_0 \approx \{ 0.028, 0.032, 0.036, 0.040 \}$.

\begin{table}[ht!]
    \centering
    \begin{tabular}{c  c  c}\hline\hline
    Parameter                           & Value in code units                                         & Comparison with mass \\\hline
    Potential coupling $\lambda$        & $1.5$                                                       & $-$ \\
    Potential barrier width $\jj_0$     & $2\pi/4.5$                                                  & $-$ \\
    Potential barrier height $V_0$      & $0.008 \jj_0 \approx 0.011$                                 & $-$ \\
    Bare field mass $m$             & $\sqrt{V_0 \jj_0^{-2} \left( \lambda^2 - 1 \right)} = 0.1$  & $m$ \\
    Parameter in the initial spectrum $T$ & $0.09, 0.1, 0.11, 0.12$                                    & $0.9m, m, 1.1m, 1.2m$ \\
    Physical lattice size $L$           & $50\sqrt{2} \jj_0 / \sqrt{V_0} \approx 791$                 & $80 / m$ \\ 
    Lattice sample points $N$           & $1024$                                                      & $-$ \\ 
    Maximum evolution time              & $5L \approx 3953$                                           & $400 / m$ \\ 
    Lattice spacing $\dd r$             & $L/N \approx 0.77$                                          & $0.08 / m$ \\ 
    Integration time step $\dd t$       & $\dd r/16 \approx 0.048$                                    & $0.0005 / m$ \\ 
    IR spectral cutoff $k_{\rm IR}$     & $\dd k = 2\pi/L \approx 7.95\times10^{-3}$                          & $0.08 m$ \\ 
    UV spectral cutoff $k_{\rm UV}$     & $\dd k N/2 \approx 4.1$                                     & $41 m$ \\
    Critical bubble energy $E_\crit$           & $\approx 0.33$                                              & $3.3 m$ \\\hline\hline
    \end{tabular}
    \caption{List of relevant physical quantities together with their numerical values and comparison with the mass scale.}
    \label{table:1}
\end{table}

\section{Observables in vacuum decay}\label{sec:ResultsObservables}

In this section, we analyze the ensembles defined above to identify new classes of observables in vacuum decay. This program was initiated in previous work~\cite{Pirvu:2021roq} where it was demonstrated that bubble nucleation centers cluster, and their two-point correlation function $\xi_{\rm bb} (t,r)$ was measured using simulations. Here, we demonstrate that a detailed study of nucleation events can reveal additional observables in vacuum decay. An overview is presented in Fig.~\ref{fig:bubble_cartoon}. We expect the qualitative results to apply to a broad set of classical excited states about the false vacuum characterized by large occupation numbers and stochastic phases of the field modes.

We begin by examining the fluctuations about the false vacuum prior to bubble nucleation and measure the effective mass and temperature that the field fluctuations evolve under. Next we measure the decay rate observed in our ensemble of lattice simulations as a function of the effective temperature and find good agreement with the theoretical prediction of Euclidean instanton theory for finite temperature vacuum decay, as reviewed in Section~\ref{sec:InstantonSection}. Moving on to study individual nucleation events, a number of features are evident. First, we demonstrate that typical bubbles do not form at rest, but rather have a center-of-mass velocity distribution that agrees well with the hypothesis that boosted bubbles are Boltzmann suppressed. By stacking many nucleation events in their rest frame, we determine the critical bubble configuration on the lattice. This empirical profile closely matches the analytical critical bubble solution. Next, we observe that bubble nucleation is preceded by a long-lived oscillon precursor field configuration. Finally, we compare three different ways of determining the critical bubble energy: from the static solution of Eq.~\eqref{eq:statsphaleron}, from the stacked decaying numerical simulations, and from the bubble velocity distribution. We find them to be in good agreement. The analysis techniques we present can be used in future lattice simulations or experiments using analogue quantum simulators of vacuum decay.

\subsection{Fluctuations around the false vacuum before decay}\label{subsec:powespec_evol}

Beyond the formation of bubbles of the true vacuum, the non-linear nature of the potential coupled with the finite size of the system and finite lattice spacing have several important implications.
\begin{itemize}
    \item The existence of field fluctuations around the minimum leads to a renormalization of the coupling constants $m$ and $\lambda$ for the effective potential seen by infrared modes on the lattice. These corrections depend on the field variance. These effects have been studied quantitatively in~\cite{braden-massreno} for the potential we use here; for a detailed discussion of renormalization effects on thermal vacuum decay see \eg~\cite{STRUMIA1999482, GLEISER199869}. 
    \item Because the system is initial out of equilibrium, the statistics of the fluctuations about the false vacuum evolve in time. In particular, power is transferred from the IR to the UV. In principle, if the false vacuum were arbitrarily long-lived, the system could thermalize to a Rayleigh-Jeans spectrum with a new temperature.  
    \item The effective couplings seen by IR modes on the lattice depend on the spectrum of fluctuations about the false vacuum. Since this spectrum is time-dependent, there will be time evolution of the properties of the IR modes which participate in vacuum decay. 
\end{itemize}
We choose the parameters of our simulations to mitigate the consequences of these effects. For the Bose-Einstein distribution with our choice of parameters, contributions to the field variance near the ultraviolet cutoff are small $\ev{\delta \jj_{k_{UV}}^2} \to 0$. In this limit, the corrections to coupling constants in the potential depend only on the amplitude of long modes with $\omega\lesssim T$. We can then choose the values of $T$ that are small enough such that corrections to the potential are small, but large enough to yield vacuum decay on a reasonable computational timescale. These choices also mitigate the second and third items above, since typical fluctuations do not experience the strong non-linearities away from the false vacuum and power that does `cascade' to the UV does not contribute significantly to the field variance. Nonetheless, it is possible to extract empirically the effective mass and temperature of the IR fluctuations around the false vacuum. Since the critical bubble is itself made up of modes with wavenumber $k \sim R_\crit^{-1}\sim m \, \mathcal{O}(1)$, the occupation numbers of the long-wavelength modes and any changes from the initial theoretical spectrum will affect the observables of vacuum decay. We investigate these changes quantitatively in this section.

\begin{figure}
    \centering
    \includegraphics[width=0.7\textwidth]{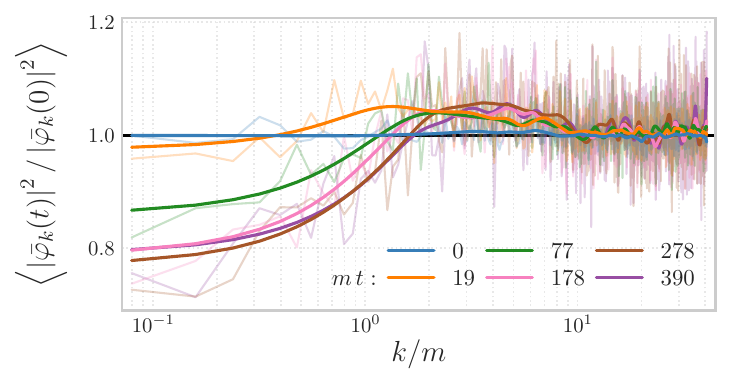}
    \caption{Ensemble averaged power spectrum at $T=1.2m$ as a function of time, normalized with respect to the theoretical average power spectrum on the initial slice, $n_k/\omega_k$ as defined in Eq.~\eqref{eq:power_spectra}. The faint lines show the true data, while the solid lines have been smoothed with a Gaussian kernel of width $0.48m$ to visualize the shape of the spectrum. The ensemble average is performed over the realizations that have not decayed by the time $t$. Since the number of surviving simulations decreases as a function of time, the statistical fluctuations on the average spectrum increase at late times. Barring this effect, there is no significant change in the UV end of the power spectrum. In the IR however, the spectral amplitude oscillates over very long timescales. In the process, the power from the largest scales $k \leq m$ migrates slowly towards the UV. This is evidence that thermalization is an extremely slow process of transferring power from the IR towards the UV, as more and more modes interact to reach a local thermal equilibrium. During this process the effective temperature is determined by the power spectral amplitude on those scales which are in thermal equilibrium.}
    \label{fig:power_spectrum_time_evolution}
\end{figure}

We begin by focusing on the members of our ensemble that do not decay. The initial conditions for these simulations are drawn from the Bose-Einstein distribution according to Eq.~\eqref{eq:power_spectra}. Choosing the undecayed members of the ensemble yields an observable selection effect on the initial power spectrum in the infrared. In simulations where the initial power on scales relevant to the formation of the critical bubble $k < k_\crit \sim R_\crit^{-1}$ is larger than the average, the probability to form a bubble is enhanced. Therefore, the undecayed realizations have a deficit in initial power at low-$k$. Simulations where the phases and initial momentum cause a time-dependent loss of power at low-$k$ are also members of the ensemble of undecayed solutions. The time-dependent loss of IR power in the field power spectrum with respect to the initial theoretical Bose-Einstein distribution is illustrated in Fig.~\ref{fig:power_spectrum_time_evolution}. This is due to a combination of two main effects: the favoring of an initial deficit of power on scales relevant for bubble formation and a phenomenon we refer to as thermalization, where the IR power `cascades' towards the ultraviolet over long time-scales as the closed system tends towards thermal equilibrium.

\begin{figure}
    \centering
    \includegraphics[width=0.7\textwidth]{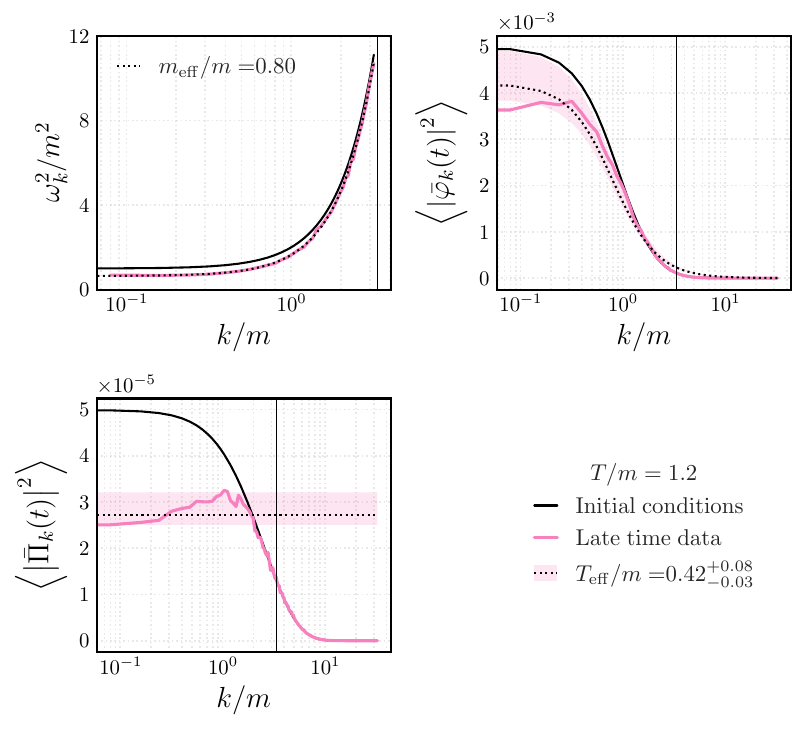}
    \caption{Measurement of the effective mass and temperature from the ensemble initialized with $T=1.2m$. Solid black curves show the dispersion relation with the bare mass in the upper left panel and the initial auto-power spectra~\eqref{eq:power_spectra} in the top right and bottom panels. The pink curves represent the ensemble average of all surviving simulations and over the time interval $160 \lesssim mt \lesssim 400$. The fit of the empirical dispersion relation with Eq.~\eqref{eq:ratio_spectra} is shown with the dotted line in the upper left panel and provides a precise determination of the effective mass. With the effective mass fixed from the dispersion relation, the field power spectrum is used to obtain a best-fit estimate of the effective temperature, assuming a Rayleigh-Jeans distribution on scales $k \sim R_{\crit}^{-1} \leq 3.3m$. The value also gives a good fit to the average power spectrum of the conjugate momentum in the same range of scales. The faded pink interval illustrates the bounds of the systematic error on the $T_\eff$ measurement. Over the time interval considered, the empirical effective temperature is constant up to statistical fluctuations. This is evidence that the modes relevant for bubble formation have reached an approximate thermal equilibrium.}
    \label{fig:thermalization_stuff}
\end{figure}

The fact that the high-$k$ part of the power spectrum remains largely unperturbed implies that the time-dependent corrections to the couplings parameters experienced by UV modes is minimal over the timescales relevant here and that the far-UV tail of our field and its conjugate momentum remain exponentially suppressed. We conclude that UV effects associated with the finite lattice spacing are not important for the analysis below. On the other hand, the changes in the IR are visible in the power spectrum, as shown in Fig.~\ref{fig:power_spectrum_time_evolution} and Fig.~\ref{fig:thermalization_stuff}. This observation signals that the effective mass and temperature controlling the power on large scales are running.

\begin{figure}[!]
    \centering
    \includegraphics[width=.9\textwidth]{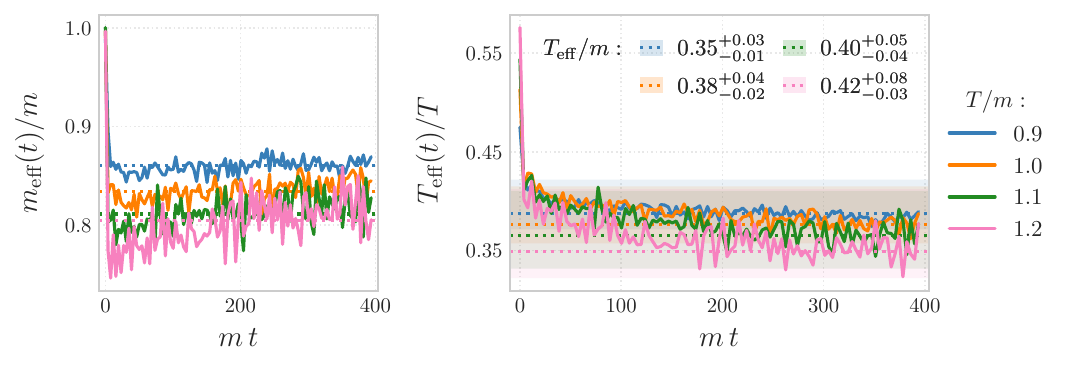}
    \caption{In the left panel we show the time-evolution of the effective mass measured by fitting Eq.~\eqref{eq:ratio_spectra} to the dispersion relation for a massive scalar $\omega_k^2 = m_{\eff}^2 + k^2$. In the right panel we show the time-evolution of the effective temperature measured by fitting the field variance to the Rayleigh-Jeans distribution with the dispersion relation fixed by the value of $m_{\rm eff}$ measured in the left panel.}
    \label{fig:meff_teff_vs_t}
\end{figure}

It is possible to measure the mass from the numerical dispersion relation using the power spectral density $\ev{ \left| \delta \jj(\omega, k)\right|^2 }$ as was done in~\cite{braden-massreno}. Here we use an alternative approach and measure the ratio between the conjugate momentum and field power spectra:
\begin{equation}\label{eq:ratio_spectra}
    \omega_k^2(t) = \frac{\ev{ \left|\delta \Pi_k(t)\right|^2 }}{ \ev{ \left|\delta \jj_k(t)\right|^2 }} = m^2_{\eff}(t) + k^2,
\end{equation}
where we explicitly indicate the $t$-dependence to stress that the power spectra and coupling constants, in particular the effective mass and temperature, are dynamical quantities. In Fig.~\ref{fig:thermalization_stuff} we show the late-time ensemble averaged power spectra of the field (top right) and conjugate momentum (bottom left) for the ensemble of simulations at $T=1.2 m$. In the top left panel we show the ratio Eq.~\eqref{eq:ratio_spectra} as well as the dispersion relation $\omega_k^2 = m_{\eff}^2 + k^2$ with the best-fit value of $m_{\eff}^2$. There is an excellent fit for $m_{\rm eff}/m = 0.80$, demonstrating that at late times the field fluctuations renormalize the mass parameter for IR modes to be {\em less} than the input value. In the left panel of Fig.~\ref{fig:meff_teff_vs_t}, we show the time-dependence of the best-fit $m_{\eff}$ at each temperature. There is a stage immediately following initialization where the mass measured from Eq.~\eqref{eq:ratio_spectra} abruptly decreases between $15\%$ and $25\%$, with the effect being stronger for larger $T$. After this stage, the mass starts to increase slowly until it reaches a plateau. For all ensembles, the effective mass reaches steady state around $m t \approx 160$. We treat the late-time value for the mass as the effective equilibrium field mass in each ensemble.

As the power `cascades' from the IR towards the UV, the field and conjugate momentum achieve a state of local thermodynamic equilibrium with an effective temperature determined by the long-wavelength modes~\cite{Boyanovsky-thermalization}. The modes which are in equilibrium will satisfy the Rayleigh-Jeans distribution defined by: $\ev{ \left|\delta \jj_k\right|^2 } \propto T_\eff/\omega_k^2, \ev{ \left|\delta \Pi_k\right|^2 } \propto T_\eff$. Note that the low-k limit of these relations implies that a decrease in the effective mass must come with a corresponding decrease in the effective temperature at fixed field variance. The measured effective temperature of the low-$k$ modes as a function of time in each ensemble is shown in the right panel of Fig.~\ref{fig:meff_teff_vs_t}. Assuming thermal equilibrium is reached on scales $k\leq 3.3m$, we use the Rayleigh-Jeans distribution to find the best-fit effective temperature from the field power spectra. In this procedure we fix the mass to the value given by our late-time measurement of the dispersion relation from the ratio Eq.~\eqref{eq:ratio_spectra}. We find that the temperature drops abruptly after initialization, reaching a constant at late times. On time scales $m t > 160$ the temperature is constant, up to statistical fluctuations due to limited sample size. We estimate the systematic error on the effective temperature measurement by looking at the spread of the conjugate momentum power spectrum on large scales, which is a direct measurement of $T_\eff$. Fig.~\ref{fig:thermalization_stuff} illustrates the measurement of the effective temperature and its error bars for the case $T=1.2 m$. We observe that $T_\eff$ is about three times lower than the `temperature' parameter $T$ in the input Bose-Einstein distribution.

For the rest of this work, we will use the values of the effective temperatures measured in this way, as well as the interpretation of local thermal equilibrium described above, to estimate the false vacuum decay rate and explain quantitatively the measured observables that we introduce.

\subsection{Decay rate}\label{subsec:decayrate}

The probability of the field remaining in the false vacuum centered at $\jj_\fv$ after a time $t$ can be parametrized as:
\begin{equation}\label{eq:decay_rate_exp}
    {\rm Pr(survive)}=e^{-\Gamma L\left(t-t_0\right)},
\end{equation}
where $\Gamma$ is the probability per unit time per unit length to form a bubble and $t_0$ is a free parameter. The instanton prediction for $\Gamma$ in Eq.~\eqref{eq:decay_rate_bounce} depends on the temperature and the critical bubble configuration. We empirically determine $\Gamma L$ by measuring the survival probability as a function of time in our ensemble of simulations. To do so, we implement a similar technique as described in previous work~\cite{braden-newsemiclassical, jenkins2023tabletop} where the survival probability is defined as the number of realizations out of an ensemble that have not nucleated a bubble by time $t$. The nucleation time is determined by the condition that the quantity $\ev{\cos{\bar{\jj}}}$, where here $\ev{\cdot}$ denotes a lattice volume average, has reached a value greater than $-0.7$. The survival fraction is fit by an exponential according to Eq.~\eqref{eq:decay_rate_exp}, and the slope of the exponent is identified with the decay rate $\Gamma L$.

\begin{figure}[t!]
    \centering
    \includegraphics[width=0.9\textwidth]{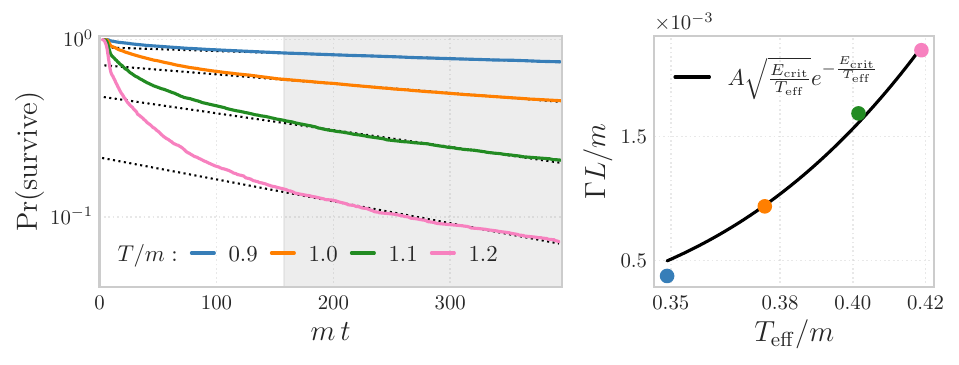}
    \caption{\emph{Left:} False vacuum survival probability as the fraction of simulations that have not produced a bubble as a function of time, estimated from ensembles of $4000$ realizations for each of the four initial power spectra differing by the value of the parameter $T$. The gray shaded region denotes the range used to fit an exponential decay rate of the form Eq.~\eqref{eq:decay_rate_exp} used to extract $\Gamma L$ in each curve. This is the time range over which the effective temperature $T_\eff(t)$ has reached a plateau (see Fig.~\ref{fig:meff_teff_vs_t}). \emph{Right:} Decay rate as a function of effective temperature, determined from the field power spectrum as explained in Section~\ref{subsec:powespec_evol}. The prediction from instanton theory is shown in black, where the prefactor $A$ has been adjusted to provide the best fit to the data and $E_\crit$ is fixed by the critical bubble configuration in the bare potential.}
    \label{fig:decay_rate_and_surv_fraction}
\end{figure}

The survival fraction for all choices of $T$ is shown in left panel of Fig.~\ref{fig:decay_rate_and_surv_fraction}. After a transient phase, the survival probability is well fitted by an exponential. We take the transient to be the time interval over which the power spectrum adjusts to the effective mass and temperature parameters, as measured in the previous section. We perform the fit over the shaded region in the figure, which excludes the transient region and encompasses three lattice crossing times, \ie for $160 < mt < 400$. We find the best-fit value from $4000$ simulations for $\Gamma L$ and $t_0$ in expression Eq.~\eqref{eq:decay_rate_exp} for each input power spectrum. In the right panel of Fig.~\ref{fig:decay_rate_and_surv_fraction} we show the trend of the decay rate with temperature compared against the instanton prediction. Specifically, the solid black line represents the best-fit curve with a functional expression given by Eq.~\eqref{eq:decay_rate_bounce}, where $B = E_\crit/T_\eff$ is fixed, and only the prefactor $A$ is a free parameter. The effective temperature is fixed by the analysis of the previous subsection. By inspection, the decay rate agrees well with the Euclidean prediction.

\subsection{Center-of-mass velocity distribution}\label{subsec:COMvelocitydistrib}

\begin{figure}[!]
    \centering
    \includegraphics[width=0.7\linewidth]{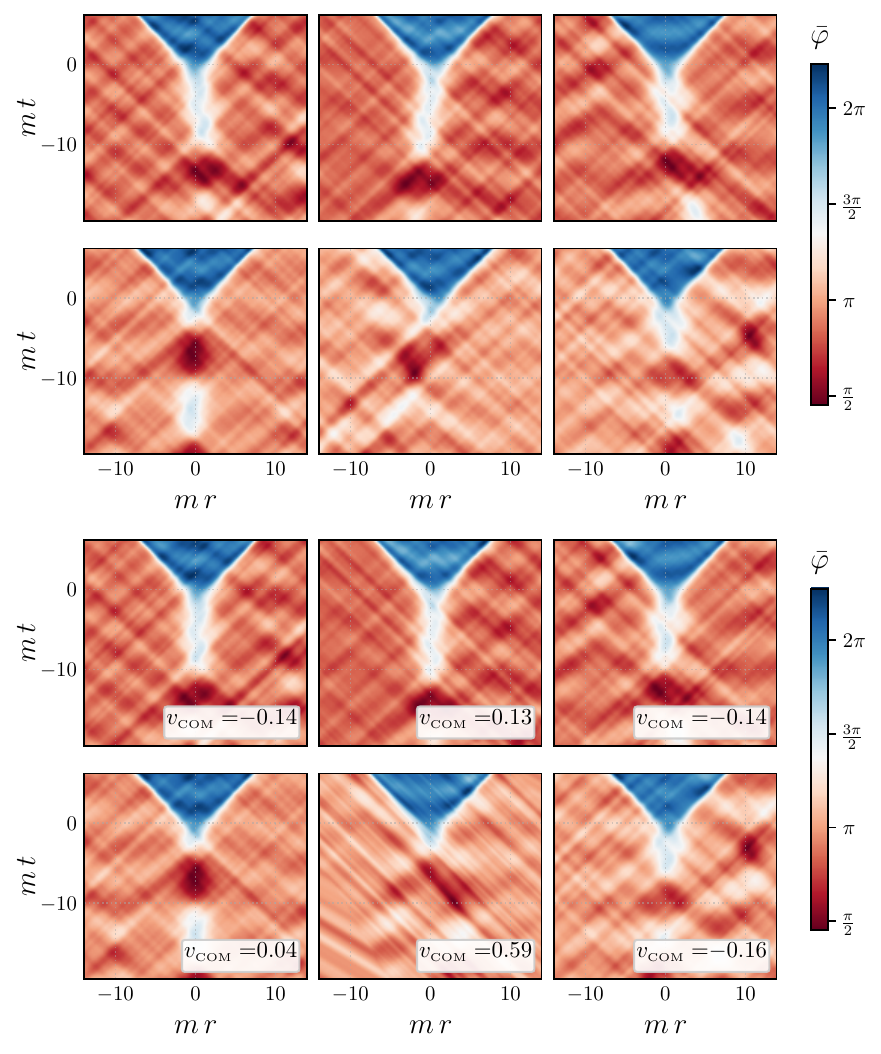}
    \caption{\emph{Top panel}: Six examples of original bubbles from the $T/m=0.9$ ensemble. \emph{Bottom panel:} The same bubbles as the top panel, after going through the de-boosting procedure. The bubble center-of-mass velocity is indicated in each case. The purpose of the Lorentz boost is to maximize the symmetry of expansion of the bubble walls.}
\label{fig:many_examples_deboost_before_after}
\end{figure}

Examining individual nucleation events in the simulations it is evident that bubbles do not nucleate at rest; several examples are shown in Fig.~\ref{fig:many_examples_deboost_before_after} and Fig.~\ref{fig:examples_deboost}. This was also observed in our previous work on bubble correlation functions~\cite{Pirvu:2021roq}. For the potential and range of temperatures studied here, we find that bubbles materialize with center-of-mass velocities ranging from $0$ up to $80\%$ the speed of light on the lattice. In this section we describe an algorithm to identify bubble nucleation events, and determine the Lorentz boost necessary to transform to the rest frame of the nucleation event.

In Section~\ref{sec:InstantonSection} we described how a critical bubble formed from the thermal ensemble is unstable to either growth or collapse. Here, we focus on nucleation events that lead to expanding bubbles. Once a critical bubble begins expanding, the surfaces of constant field describing the bubble walls asymptote to timelike hyperbolas (studied quantitatively in Ref.~\cite{Aguirre:2008wy}). We first identify simulations where the field value achieves this for one of the two true vacua at $\bar{\jj}_\tv = 0$ and $\bar{\jj}_\tv = 2\pi$. Because our potential is symmetric about the false vacuum, transitions to either of these true vacua are identical for our purposes. For realizations where bubbles nucleate to $\bar{\jj}_\tv = 0$, we reflect about the mean field and momentum values to produce a nucleation event to the true vacuum at $\bar{\jj}_\tv = 2\pi$. This doubles our sample size. In cases where we find more than one nucleation event, we truncate the simulations to encompass spacetime regions containing only the earliest expanding bubble.

From this sample, we follow a procedure similar to the one proposed in Ref.~\cite{Pirvu:2021roq} to identify bubble nucleation centers and the bubble center-of-mass velocities. The technique is based on finding the Lorentz boost that produces hyperbolic bubble walls with symmetric expansion away from a common reference point. Such symmetric expansion is what one expects to observe in the bubble's rest frame (center-of-mass frame) in absence of fluctuations. Any deviation from this symmetry indicates there is a preferred direction for the expansion, sourced by a center-of-mass velocity component. The total velocity needed to bring the bubble from the initial frame of nucleation into its rest frame via a Lorentz boost is its center-of-mass velocity $v_\com$. The sign of the velocity indicates whether the expanding bubble is moving to the left or to the right in the lattice frame. To estimate it, we define a measure for the asymmetry between the expansions of the left- and right-traveling walls. In its rest-frame, the bubble is fully symmetric, so the goal is to treat this asymmetric expansion as a residual and minimize it. The full details of the procedure can be found in Appendix~\ref{app:BB}.

Out of the four ensembles of $4000$ realizations at each value of $ T/m \in \left\{ 0.9, 1, 1.1, 1.2 \right\}$, we detected a total of $\{1003, 2192, 3165, 3711\}$ bubbles, respectively. We exclude bubbles that formed before $mt \approx 80$, to allow for thermal state to be reached. This filters out a large fraction of simulations, especially at high values of $T$. As discussed earlier, the effective temperature plateaus after $mt \geq 160$, however we keep a looser cutoff here to gain statistical power; we do not expect this choice to significantly affect our results. For each realization, we checked visually that de-boosted bubbles from the procedure described above appeared symmetric against a central axis. These requirements leave $\{559, 956, 997, 591 \}$ bubbles at rest, which made up the ensembles considered throughout the rest of this work. Out of these,
in the simulation frame $\{ 49.91\%, 51.15\%, 51.65\%, 49.41\% \}$ were right-movers, and the rest were left-movers -- a nearly even distribution as expected. In absolute value, $\{31.48\%, 33.58\%, 38.72\%, 42.81\%\}$ were moving faster than $v_\com=0.3$, showing a clear increase with temperature. This trend continues with a larger velocity threshold, as $\{7.69\%, 10.98\%, 13.94\%, 16.58\%\}$ of all bubbles were moving faster than half the speed of light. 

\begin{figure}[b!]
    \centering
    \includegraphics[width=.85\textwidth]{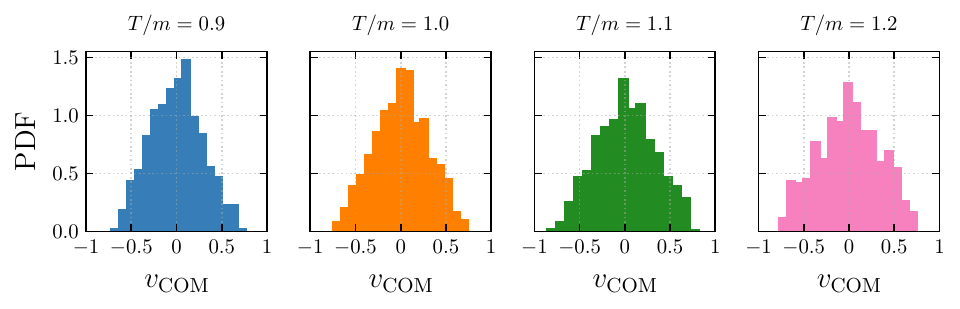}
    \caption{The normalized distributions for the measured center-of-mass velocities.}
    \label{fig:vcom_distrib}
\end{figure}

\begin{figure}[t!]
    \centering
    \includegraphics[width=0.4\textwidth]{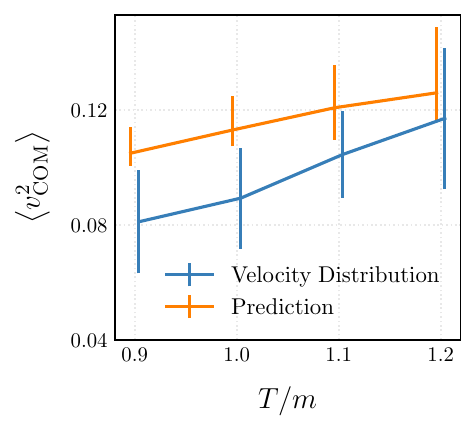}
    \caption{Comparison between the measured variance in the distribution of velocities depicted in Fig.~\ref{fig:vcom_distrib} versus the prediction in Eq.~\eqref{eq:varvel_pred} where $E_\crit$ is the energy of the critical solution from instanton theory and $T_\eff$ is given by the effective temperatures at late time in each ensemble. There is good agreement between the two.}
    \label{fig:comparison_vcom_pred}
\end{figure}

The full distribution of velocities found in each ensemble is shown in Fig.~\ref{fig:vcom_distrib} for each initial power spectrum. The mean of these distributions is near zero; the variance is plotted in Fig.~\ref{fig:comparison_vcom_pred}. To obtain an estimate for the magnitude of the error in our result, we divided each ensemble into $15$ sub-ensembles and computed the variance. Then we used the standard deviation of the resulting distribution as an estimate for the error.

This distribution is described by the following simple theoretical model. A boosted bubble has total energy 
\begin{equation}\label{eq:Ebubble}
    E=\gamma E_\crit\approx E_\crit (1+v_\com^2/2), 
\end{equation}
where in the second equality we have replaced the boost factor $\gamma = (1-v_\com^2)^{-1/2}$ by its non-relativistic approximation which is accurate for the majority of the bubbles used in our analysis. In a thermal ensemble with temperature $T_\eff$ the probability to find such bubble must obey the Boltzmann distribution,
\begin{equation}\label{eq:pgamma}
    P(v_\com) = \mathcal{N}^{-1} \exp{ -\frac{v_\com^2 E_\crit}{2T_\eff} },
\end{equation}
with the normalization factor\footnote{We formally extended the range of integration in the normalization condition $\int \dd v P(v)=1$ from $-\infty$ to $+\infty$, which is justified if $E_\crit\gg T_\eff$, so that the distribution is peaked at $v \ll 1$.} $\mathcal{N} = \sqrt{2\pi T / E}$. The expectation value for the variance is:
\begin{equation}\label{eq:varvel_pred}
    \ev{ v_\com^2 } =\int \dd v \,v^2 P(v)= \frac{T_\eff}{E_\crit}.
\end{equation}
Note that this expression coincides with the inverse critical bubble action~\eqref{eq:Bbubble}, $\ev{ v_\com^2 } = B^{-1}$. Using the energy of the critical bubble obtained by solving Eq.~\eqref{eq:statsphaleron} and the empirical values for the effective temperature, this expression gives us a theoretical prediction for the center-of-mass velocity. This is compared with the variance measured in the simulations in Fig.~\ref{fig:comparison_vcom_pred}. The agreement is within one sigma of the empirically determined variance.

\begin{figure}[h!]
    \centering
    \includegraphics[width=0.5\textwidth]{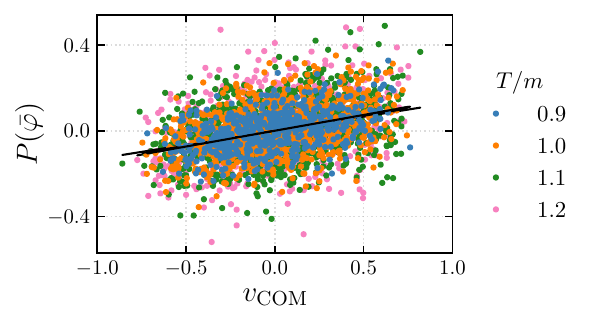}
    \caption{A scatter plot of the total conserved relativistic momentum on the lattice at initialization versus bubble center-of-mass velocity. For reference, the initial variance of fluctuations in each ensemble was $\sigma_{\bar{\jj}}\approx \{ 0.27, 0.30, 0.33, 0.36\}$ respectively. There are four solid black lines that represent the best linear fit through the data from each ensemble. They all share the same slope to within the thickness of the lines, showing that the degree of correlation remains constant with temperature. The root-mean-square in momentum increases as a function of $T$, as expected.}
    \label{fig:initial_conditions}
\end{figure}

The relativistic Klein-Gordon field has two associated conserved charges: the total energy and the total momentum. These remain conserved to near machine precision over the entire time of the evolution. They are defined on the lattice as:
\begin{equation}\label{eq:Noether_energy}
    H(\bar{\jj}) = \sum_{r}^L \left[\frac{1}{2} \bar{\Pi}^2+\frac{1}{2}|\partial_{r} \bar{\jj}|^2 + V(\bar{\jj}) \right] = H(\jj) / \jj_0^2,,
\end{equation}
\begin{equation}\label{eq:Noether_momentum}
    P(\bar{\jj}) = - \sum_{r}^L \bar{\Pi} \partial_{r} \bar{\jj} = P(\jj) / \jj_0^2.
\end{equation}
Both quantities are fixed by the initial conditions. However, different realizations have a spread in the initial energy and momentum due to the stochastic sampling of the field and momentum mode amplitudes. In Fig.~\ref{fig:initial_conditions} we plot the initial relativistic momentum defined in Eq.~\eqref{eq:Noether_momentum} versus the measured center-of-mass velocity $v_\com$ for each realization. The best-fit linear correlation between the total momentum and measured center-of-mass velocity is nearly the same in each ensemble. The spread about the mean correlation increases with $T$, as expected from the expression for the variance in initial momenta (see Eq.~\eqref{eq:initialspectra}). This indicates that the initial conditions influence the dynamics of the bubble at nucleation. The average relativistic momentum is zero across the ensemble. However, the local surplus of momentum associated with the random initial conditions in a given realization selects a preferred frame of reference for bubble nucleation.

\subsection{Average bubble}\label{subsec:averagebubble}

In this section, we define an ensemble-averaged bubble measured on the lattice and compare it to the critical bubble predicted by the thermal Euclidean instanton described in Section~\ref{sec:InstantonSection}. A non-zero center-of-mass velocity at nucleation leads to morphological changes in the critical bubble; for example, its size at nucleation is length-contracted, while the total energy is increased. Therefore, we must first transform to the frame where bubbles are at rest as, described above. Even after boosting to the rest frame, there is still great diversity in the details of individual nucleation events. For example, in Fig.~\ref{fig:many_examples_deboost_before_after}, some bubbles loiter around the turnover value of the potential, while others make the transition from the false vacuum to the true vacuum much faster. The variance between different realizations is greatest near the nucleation center, where one would like to make a direct comparison with the critical bubble solution. Here, we define a stacking procedure to compute the ensemble-averaged bubble. Our algorithm is fully automated and does not use any prior information about the expected profile.

The main idea is based on the observation that the walls of a stationary bubble start at rest at nucleation, then expand with acceleration, asymptoting to $c=1$. As they accelerate, the walls undergo Lorentz contraction, becoming thinner, and gaining a momentum far greater than the typical momenta in the fluctuations around the false vacuum. Therefore, relativistic walls become insensitive to fluctuations about the false vacuum and expand at the same rate across all realizations. Essentially, bubbles have different formation histories, but expand in a universal manner at late times.

To stack the bubbles, we need a reference point that is common to all cases. We call this reference point the spacetime location of the bubble's `nucleation' and label its coordinates as $(r_{\N}, t_{\N})$. These coordinates are different for every bubble in the ensemble. To determine $t_{\N}$, we search for the time slice where the bubble has reached a radius $R$ where the field amplitude is above a specific threshold $\bar{\phi}$. In other words, the bubble is identified as the region of width $2R$ where the field amplitude is greater than $\bar{\phi}$. We also need to define $r_{\N}$ for each simulation. Since all bubbles have been already de-boosted and are assumed to be at rest, we make use of their symmetry to define the nucleation center as the midpoint of this region. The bubbles are translated to grids centered at $(r_{\N}, t_{\N})$, and the stack average is computed with respect to this consistently defined grid.

\begin{figure}[t!]
    \centering
    \includegraphics[width=0.4\textwidth]{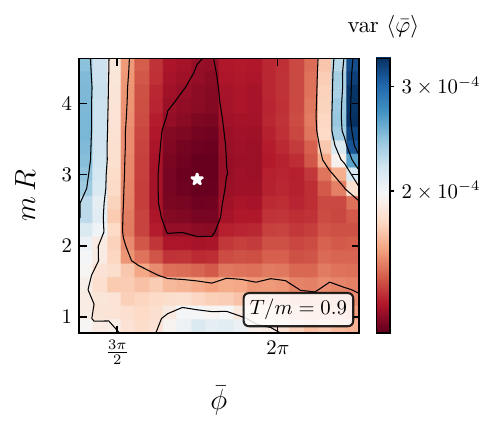}
    \caption{Sample variance (defined in Eq.~\eqref{eq:sample_var}) around the nucleation region of the stacked average bubble for $T=0.9m$. Each point on the graph shows the value of the variance as a function of the field amplitude threshold $\bar{\phi}$ and bubble size $R$. These two parameters control where the coordinates of the nucleation center are assigned in each simulation. The white star shows the location where the minimum variance is achieved. Each ensemble yields a different combination that satisfies this condition. The average bubble is produced by stacking bubble nucleation events defined by the $(\bar{\phi},R)$ pair at this point.}
    \label{fig:sample_variance_on_average_bubble}
\end{figure}

To find the best choice for $\bar{\phi}$ and $R$, we scanned over a wide and physically motivated range of values. We choose $mR$ between $0.77$ (slightly smaller than the instanton prediction) and $4.63$ (roughly three times the instanton prediction). For $\bar{\phi}$, we choose a range between $\bar{\jj}_\fv + \sigma_{\bar{\jj}}$ (far lower amplitude than expected from the instanton prediction) and $\bar{\jj}_\fv + 6\sigma_{\bar{\jj}}$ (far larger in amplitude than the instanton prediction). To estimate the goodness of fit, we minimize the sample variance over a finite spacetime region centered at the nucleation center. The sample variance is computed as
\begin{equation}\label{eq:sample_var}
    \mathrm{var} \left< \bar{\jj} \right> = \frac{1}{
    \left( 2 \Delta r \right)^2 } \sum_{r,t} \Big|\frac{1}{S} \sum_{i=1}^S \left[ \bar{\jj}_i(r,t) - \left< \bar{\jj}(r,t) \right> \right]^2 \Big|,
\end{equation}
where $S$ is the total number of samples in the stack. The variance is computed for every pair of field amplitude value $\bar{\phi}$ and bubble width $R$ considered. The nucleation region is taken to be the spacetime volume defined by $r \in \left[r_{\N} - \Delta r, r_{\N} + \Delta r \right]$ and $t \in \left[t_{\N} - \Delta r, t_{\N} + \Delta r\right]$ with $\Delta r = 30\dd r$ around the nucleation site at coordinates $(r_{\N}, t_{\N})$ uniquely defined by the pair $\left(\bar{\phi}, R\right)$. For reference, in mass units $m \Delta r \approx 2.4$. 
The sample variance for all combinations of $R$ and $\bar{\phi}$ in the case of the $T=0.9m$ ensemble is shown in Fig.~\ref{fig:sample_variance_on_average_bubble}. The variance is largest around the boundaries where the parameters take un-physical values. The white star denotes the combination of parameters that minimizes the variance and has been used to obtain the average bubble. This pair is different for each choice of $T$.

\begin{figure}[t!]
    \centering
    \includegraphics[width=1.\textwidth]{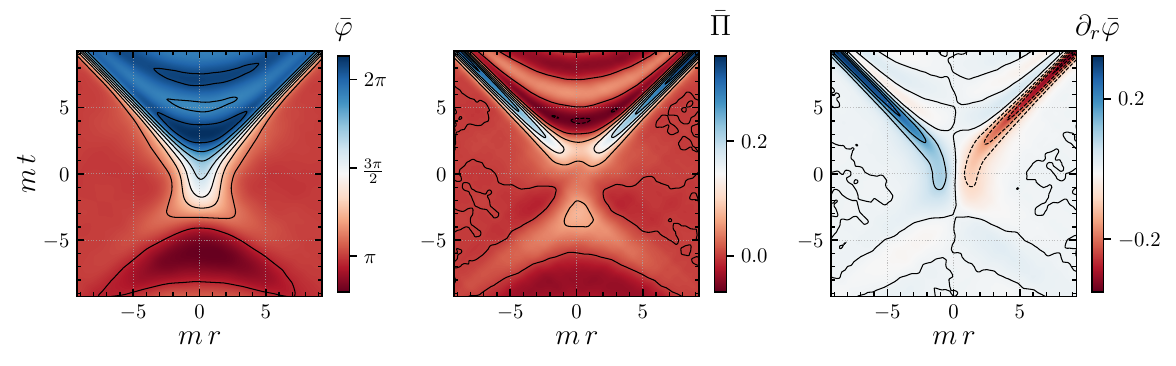}
    \caption{From left to right, we show the average bubble, average conjugate momentum and average bubble gradient for $T = 0.9 m$ obtained by stacking individual bubbles at rest. The black contours are added to help visualize surfaces of constant field amplitude. Qualitatively, beyond the nucleation time shown here at $t=0$, the features of the three fields match the characteristics of the critical saddle point solution pictured in the left panel of Fig.~\ref{fig:bare_sphaleron_and_subcritical}. Moreover, the average fields look similar, up to sample variance, among all four ensembles with different values of $T$.}
    \label{fig:average_fields}
\end{figure}

The average bubble is shown in Fig.~\ref{fig:average_fields} alongside its momentum and gradient on equivalent coordinate grids. The momentum and gradient fields have been obtained in the same manner as the procedure applied to the field. Namely, the Lorentz boosts were performed using the same coordinate grid definitions and gamma factors in each simulation, and the stacking was done with respect to the same set of reference points $(r_{\N}, t_{\N})$. Notice that the bubble profile has common features with the canonical result shown in the left-hand panel of Fig.~\ref{fig:bare_sphaleron_and_subcritical}. In particular, after a region of space where the field briefly loiters around the potential maximum, expanding bubble walls form, and the bubble interior rolls down to the true vacuum at $\bar{\jj}_\tv = 2 \pi$. The wall trajectories are most visible in the gradient plot, where it can be seen that they quickly asymptote to null and that the gradient increases as the walls length-contract while achieving increasingly high velocity. From the momentum plot, we see that the magnitude of momentum remains small until the bubble is well-formed. Finally, in the field plot, note that prior to the bubble nucleation event, there is a coherent field configuration with $\bar{\jj} < \bar{\jj}_\tv$, which is in the opposite direction in field space than the false vacuum. We will discuss this bubble precursor in the next section.

To make a comparison with the Euclidean instanton prediction for the critical bubble profile, we must select the corresponding time slice from the ensemble-averaged two-dimensional bubble shown in Fig.~\ref{fig:average_fields}. To find it, we use the following method. We sample the field time-slices around the region identified as $t=0$ in the plot. At each time step near $t \approx 0$, we take the field configuration $\bar{\jj}(r,t)$ and time-evolve it using the bare equations of motion, and assuming no thermal fluctuations. We define the critical bubble profile $\bar{\jj}_\crit(r,t_\crit)$ as the earliest time slice that evolves into an expanding bubble, in the absence of fluctuations and with zero initial momentum everywhere.

\begin{figure}
    \centering
    \includegraphics[width=0.75\textwidth]{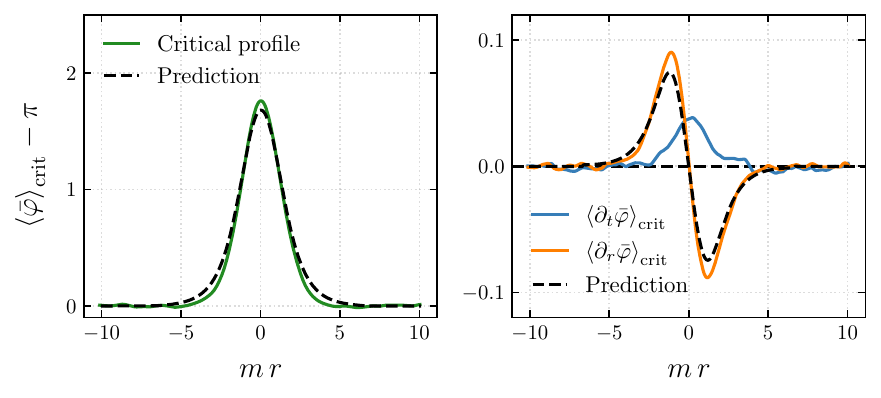}
    \caption{ \emph{Left:} The critical bubble configuration $\bar{\jj}_\crit$ is shown in solid green. This profile represents the field configuration on the critical time slice $t_{\crit}$, taken from the average bubble illustrated in Fig.~\ref{fig:average_fields}. Using this configuration as initial conditions and time-evolving under the bare equations of motion, we obtain the left panel of Fig.~\ref{fig:average_critical_solution_and_precursor}. Comparing the critical profile obtained from simulations to the Euclidean prediction, shown in dashed black, we find excellent agreement. The Euclidean profile also represents the initial conditions for the field configuration as a function of lattice site $\bar{\jj}(r,t=0)$, which was used to obtain the left-hand panel in Fig.~\ref{fig:bare_sphaleron_and_subcritical}. \emph{Right:} The average gradient and momentum on the critical time slice are compared to their theoretical predictions: the gradient of the Euclidean solution and the uniformly null momentum, respectively.}
    \label{fig:critical_profile_comparison}
\end{figure}

\begin{figure}
    \centering
    \includegraphics[width=0.8\textwidth]{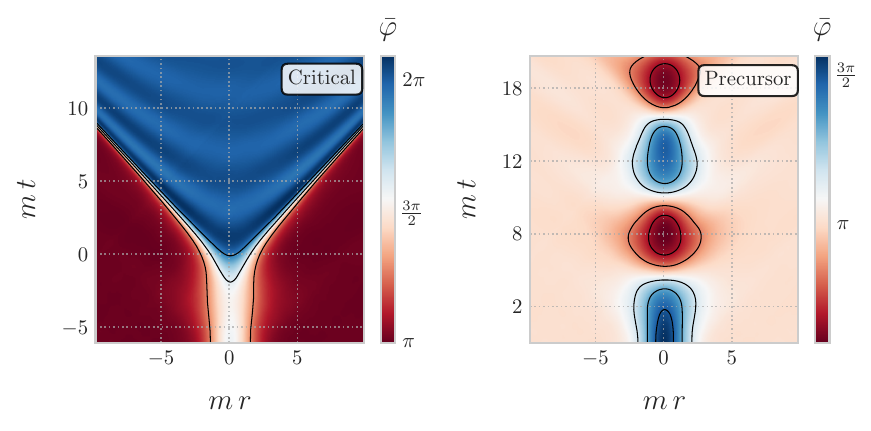}
    \caption{On the left we show the time evolution in the absence of field or momentum fluctuations of the empirical critical profile depicted in Fig.~\ref{fig:critical_profile_comparison}. On the right, we plot the time evolution of the field configuration corresponding to $\bar{\jj}(r, t_{\crit}-\dd t_{\rm out})$. The resemblance with Fig.~\ref{fig:bare_sphaleron_and_subcritical} using the Euclidean critical bubble as initial conditions is striking.}
    \label{fig:average_critical_solution_and_precursor}
\end{figure}

The profile for the $T/m=0.9$ case is shown in Fig.~\ref{fig:critical_profile_comparison}, alongside the solution found using the Euclidean instanton. The two agree quite well. The time evolution of the empirical critical bubble profile is shown in the left panel of Fig.~\ref{fig:average_critical_solution_and_precursor}. After a brief loitering period, the configuration develops into an expanding bubble. The critical bubble defined in this way is nearly identical across all four ensembles with different $T$. Additionally, using the average field profile selected from one time step earlier, at $t=t_\crit-\dd t_{\rm out}$, as an initial conditions in our code, yields the oscillon configuration depicted in the right panel of Fig.~\ref{fig:average_critical_solution_and_precursor}. Notice the striking resemblance between this and Fig.~\ref{fig:bare_sphaleron_and_subcritical} -- the ensemble-averaged sub-critical bubble from lattice simulations matches well with the expectation from the thermal Euclidean instanton.

\begin{figure}
    \centering
    \includegraphics[width=1\textwidth]{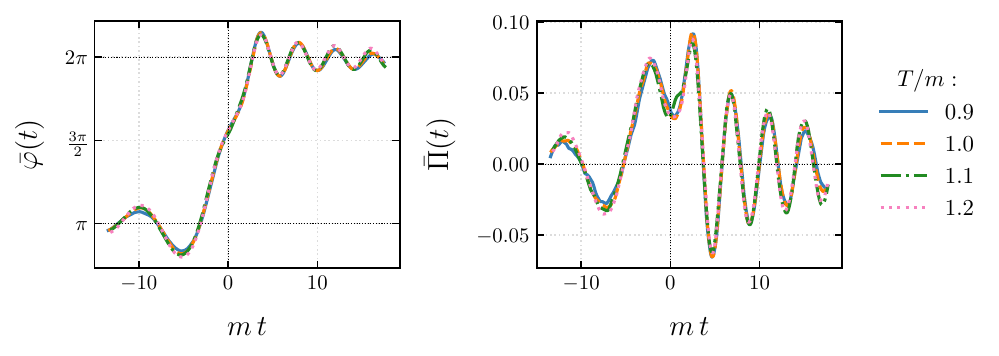}
    \caption{The time evolution of the field amplitude $\bar{\jj}$ (on the left) and the momentum amplitude $\bar{\Pi}$ (on the right) along the central axis, where $r=0$, of the average configurations depicted in Fig.~\ref{fig:average_fields}. For each ensemble labeled by the value of $T$, we identified the critical time $t_{\crit}$ as the earliest time slice where the field profile time-evolves into the expanding bubble. We offset the time evolution so that in all cases $t_{\crit}=0$. The four profiles are similar, even though they are obtained by averaging different ensembles. Before the critical time, the average field oscillates around $\bar{\jj}_\fv$, while afterwards it oscillates around $\bar{\jj}_\tv$ with decreasing amplitude as it settles into equilibrium. Around $t_{\crit}$, the field amplitude makes a jump and simultaneously the momentum temporarily acquires a non-zero amplitude, of roughly $\sigma_{\bar{\Pi}}$ in magnitude. At the critical time $t=0$, the momentum one-point function is at a local minimum.}
    \label{fig:evol_xn_vs_t}
\end{figure}

In Fig.~\ref{fig:evol_xn_vs_t}, we show the time evolution of the average field and average momentum from Fig.~\ref{fig:average_fields} for all values of $T$, at a fixed spatial coordinate $r=0$, which is the location of the nucleation center. The critical time has been identified for each of the four ensembles using the method described above. The curves in Fig.~\ref{fig:average_fields} have been matched so that $t=0$ corresponds to $t_\crit(T)$. First, note that the average evolution around this time is nearly identical for all values of $T$, although each curve was formed by averaging a completely different ensemble of bubble nucleation events. The field first oscillates about the true vacuum at $\bar{\jj}_\fv = \pi$, transitions across the potential barrier, and then oscillates about the false vacuum at $\bar{\jj}_\tv = 2\pi$. Around $mt=0$, there is a local minimum in the field momentum, as it briefly loiters around $\bar{\jj} \simeq 3\pi/2$.


However, note from the right panels in Figs.~\ref{fig:critical_profile_comparison} and~\ref{fig:evol_xn_vs_t}, as well as from the middle panel of Fig.~\ref{fig:average_fields}, that the average momentum around the critical nucleation time $t_{\crit}$ does not vanish across the lattice. This is in contrast to the Euclidean instanton solution, where the momentum vanishes everywhere. The extreme case where decays are driven entirely by a momentum profile was studied in~\cite{Blanco-Pillado_2019}. Here, we do not find that this is the dominant channel, and furthermore, the critical bubble solution we have identified does not require initial momentum to produce an expanding bubble. Since the amplitude of momentum at $(r=0, t=t_{\crit})$ from both Fig.~\ref{fig:critical_profile_comparison} and Fig.~\ref{fig:evol_xn_vs_t} is of order the average root-mean-square of momentum fluctuations in the initial conditions, $\sigma_{\bar{\Pi}} \approx \{ 0.040, 0.045, 0.051, 0.056\}$, it is possible that this is a residual of the de-boosting procedure or even a bias owing to the fact that we only average critical profiles which result in expanding bubbles, and neglect solutions that collapse back into the false vacuum via oscillons. Nevertheless, it is clear that field dynamics play an important role in bubble nucleation since the average bubble includes precursor fluctuations. We now turn to study these precursors in more detail.

\subsection{Bubble precursors}\label{subsec:Precursors}

The Euclidean instanton formalism provides a prediction for the critical bubble profile, but offers no guidance into how this configuration comes about from an ensemble of fluctuations around the false vacuum. General theoretical considerations can yield some insight, as described in Ref.~\cite{Aguirre:2011ac}. In short, the most probable formation history of a rare configuration from a thermal ensemble is the time-reverse of its decay. For thermal vacuum decay, the critical bubble is in unstable equilibrium between expansion and collapse. As we demonstrated in Fig.~\ref{fig:average_critical_solution_and_precursor}, time-evolving the slightly subcritical average bubble yields an oscillon (a stable and compact oscillating field configuration), which after long times would decay back to un-bound plane-wave fluctuations about the false vacuum. Time-reversing, the prediction is that the most probable formation history of an expanding thermal bubble is for plane waves about the false vacuum to scatter, producing an oscillon, which propagates for a long time, eventually interacting with thermal fluctuations to produce a critical expanding bubble. Indeed, oscillon precursors have been observed previously in lattice simulations~\cite{PhysRevLett.94.151601}, where they were shown to enhance the decay rate in a quench.

\begin{figure}[!]
    \centering
    \includegraphics[width=0.9\textwidth]{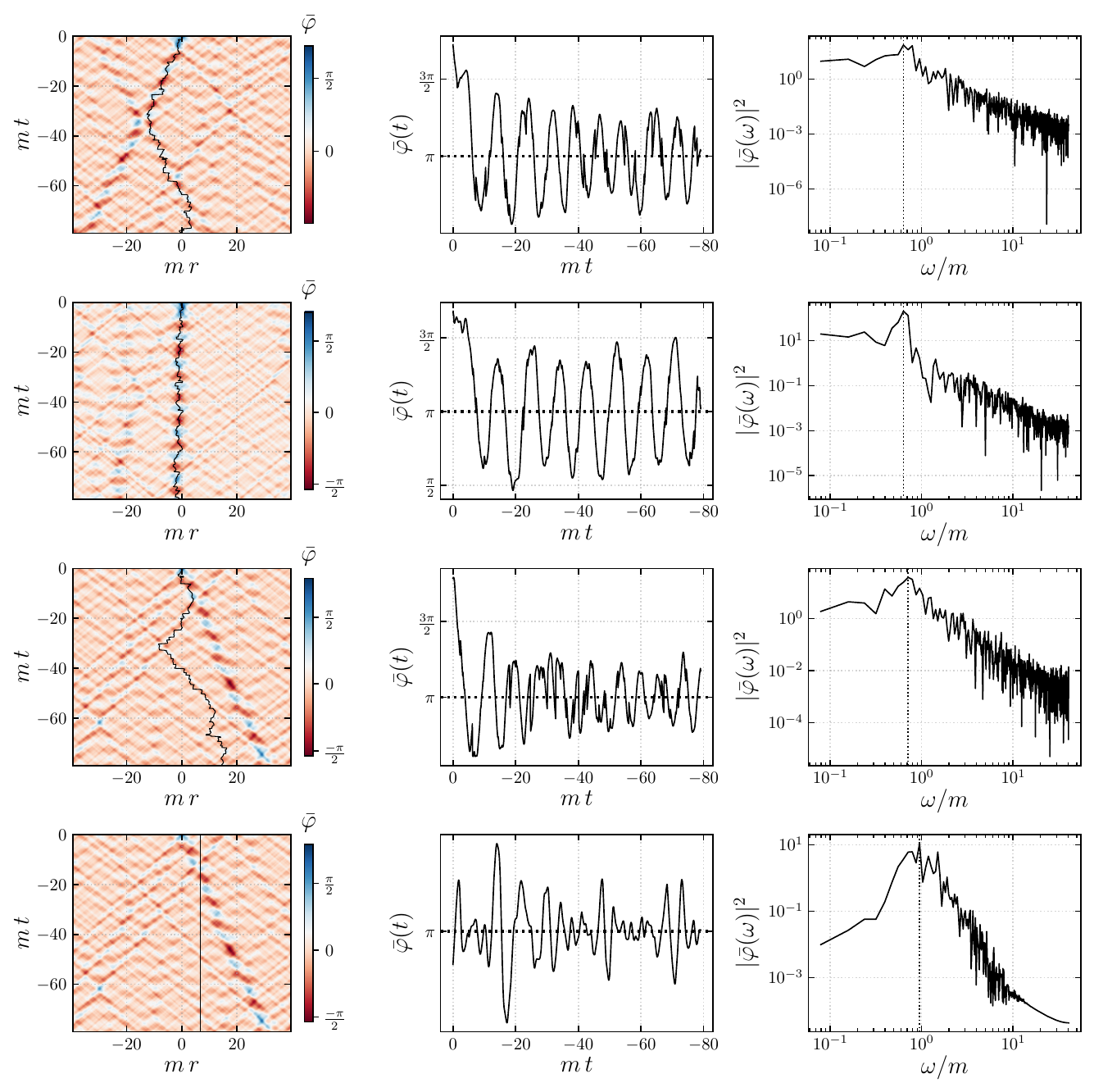}
    \caption{\emph{Left:} Examples of field configurations preceding the bubble nucleation at $(r,t)=(0,0)$. The trajectories of the oscillon precursors are shown in black. The precursors are not the only structures evolving on the lattice, and in fact in many cases it is collisions of such objects that trigger bubble nucleation. The oscillon collisions make it difficult, in general, to track the long-term evolution of the precursor, as exemplified in the third row where the trajectory seems to diverge from the dominant oscillon in that simulation at around $mt=-20$ and instead random background fluctuations are being picked up. The bottom row shows the same realization as above, but instead of the oscillon, we analyze an example null trajectory. \emph{Middle:} The field amplitude along the oscillon trajectory as a function of time. A large amplitude sinusoidal stands out in the top two cases. The bottom case exemplifies that the typical background field does not, in general, show the same phase coherence. \emph{Right:} Taking the time-domain Fourier transform of the field shown in the middle panel, we compute its power spectrum. The peak frequency $\omega_{\osc}$ is identified by the vertical dotted line. Interestingly, the frequency corresponding to this peak is below the mass scale of the field everywhere, except in the null test case. This demonstrates that the oscillons are bound structures characterized by a lower energy per particle, $\omega_{\osc} < m_\eff$, than a collection of free waves with $\omega \geq m_\eff$. Moreover, since the critical bubble is a static solution, it has $\omega_{\crit} \to 0$. In this sense, oscillons are an intermediate state between propagating field degrees of freedom and the critical bubble solution.}
    \label{fig:oscillon_examples}
\end{figure}

We can test the hypothesis that the most likely bubble formation history starts with an oscillon by using our lattice simulations to empirically measure the dynamics prior to bubble nucleation. In Fig.~\ref{fig:average_fields} we see that the average bubble configuration has a large under-density in field space, followed by a peak in momentum. Looking closely at each realization, empirically we observe that many bubbles form from oscillons; several examples are shown in Fig.~\ref{fig:oscillon_examples}.

We now explain how to disentangle these structures from the background field fluctuations. First we select the simulations where the bubble nucleates after at least a duration $mt = 80$, equivalent to a full lattice crossing time. This allows us to trace the long-term evolution of the precursors. Throughout this section we limit our discussion to the case $T/m=0.9$ where the field fluctuations are smallest, but the results presented below apply also to the other ensembles.

Oscillons stand out as large amplitude long-wavelength fluctuations. As they evolve slowly in time, they bounce around the lattice subject to a random Brownian-like motion. To consistently isolate their trajectory across realizations, we compute the instantaneous amplitude (or dynamical envelope) of the field $\bar{\jj}(r,t)$~\cite{MDFT07}. We first remove the field average over the spacetime region of interest to ensure $\ev{\bar{\jj}(r,t)} = 0$. Then, we perform a Fourier transform in time, set the negative-frequency modes to zero, and double the positive-frequency amplitudes. The inverse Fourier transform of this modified spectrum yields a complex field, whose absolute value is the envelope. The trajectory is simply given by the location $r$ on the lattice where the envelope has peaked in amplitude. The starting point for the procedure is the location of the bubble nucleation, shown as $(r,t)=(0,0)$ in the figures on the left panel of Fig.~\ref{fig:oscillon_examples}. Then we trace the trajectory backwards in time, imposing the additional requirement that the maximum at $t$ should not be farther than $mr=2$ away from the maximum computed at $t+\dd t_{\rm out}$. The time evolution is truncated at $mt = -80$. In the middle panels of this figure, we plot the time-development of the field along this trajectory. In the right panels, we show the spectral content of the field along this trajectory.

\begin{figure}[!]
    \centering
    \includegraphics[width=1\textwidth]{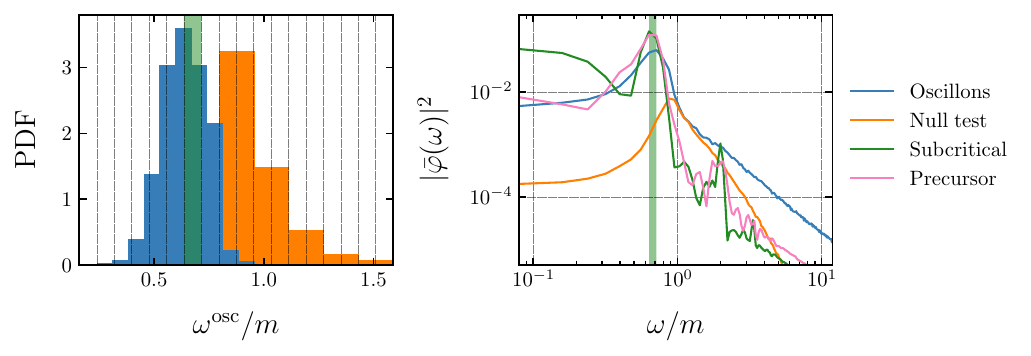}
    \caption{\emph{Left:} In blue is the distribution of characteristic frequencies of the bubble precursors $\omega_\osc$. In orange is the distribution of the peak frequencies in the power spectrum of null field trajectories drawn from the same realizations. The oscillon distribution has smaller variance and peaks at lower frequency than the null trajectories. This is consistent with the bubble precursors being bound states of similar properties across the ensemble. \emph{Right:} The average oscillon (in blue) and the average null trajectory (orange) power spectra. We also show in green the spectrum of the $\bar{\jj}_{\crit}(r=0,t)$ trajectory of the precursor in Fig.~\ref{fig:bare_sphaleron_and_subcritical} and in pink the subcritical average bubble solution from Fig.~\ref{fig:average_critical_solution_and_precursor}, for comparison. With the exception of the null test line which peaks around $\omega \approx m_\eff$, all other spectra peak over the same frequency range, highlighted by the faded green band in both images. The occupation numbers are also much larger on large scales for the three spectra that measure oscillons, which supports their identification as bubble precursors.}
    \label{fig:main_frequency_of_osc_comparison}
\end{figure}

Notice from the middle and right panels of Fig.~\ref{fig:oscillon_examples} that the field along the trajectory oscillates with a characteristic oscillon frequency $\omega_\osc$. This stands out as a peak in the power spectrum. We associate a characteristic $\omega_\osc$ to each bubble precursor in our ensemble. Their distribution is shown in blue in the left panel of Fig.~\ref{fig:main_frequency_of_osc_comparison}. To highlight the difference between oscillons and the background fluctuations, we pick random field trajectories through each simulation and select the frequency where their respective power spectra peak. This distribution is plotted in orange. Note that the peak of the power spectrum for oscillon trajectories is lower than the peak on random trajectories (appropriately centered on $\omega = m_\eff$, the prediction from the dispersion relation for plane waves about the false vacuum). This is consistent with the interpretation that oscillons are bound states.

Next, we compute the average power spectrum of all oscillons and the average power spectrum of random trajectories and plot them side by side in the right panel of Fig.~\ref{fig:main_frequency_of_osc_comparison}. We also plot, for comparison, the power spectra of the field trajectories $\bar{\jj}_\crit(r=0, t)$ for both the subcritical bare lattice solution in Fig.~\ref{fig:bare_sphaleron_and_subcritical} and the precursor to the average bubble critical solution shown in Fig.~\ref{fig:average_critical_solution_and_precursor}. These are shown in green and pink, respectively. Notice that these peak at the same frequency as the average oscillon signal. This is highlighted by the green band in both plots. This is strong evidence that the formation history of bubbles in our simulations includes an oscillon precursor matching the oscillon that results from collapse of a subcritical bubble (as in e.g. Fig.~\ref{fig:bare_sphaleron_and_subcritical}).

Here we considered the trajectory of the oscillons in the original simulations, before de-boosting. In principle the distribution of velocities for the oscillons should obey $\ev{v_{\osc}^2} \propto T_\eff/E_\osc$ but the frequent collisions make it difficult to systematically measure their velocities (such a collision which changes the direction of motion of the oscillon is clearly visible in the top panel in Fig.~\ref{fig:oscillon_examples}). We will perform a more detailed study of the properties of oscillon precursors in future work.

\subsection{Critical bubble energy}\label{subsec:critical_energy}

\begin{figure}[t!]
    \centering
    \includegraphics[width=0.7\textwidth]{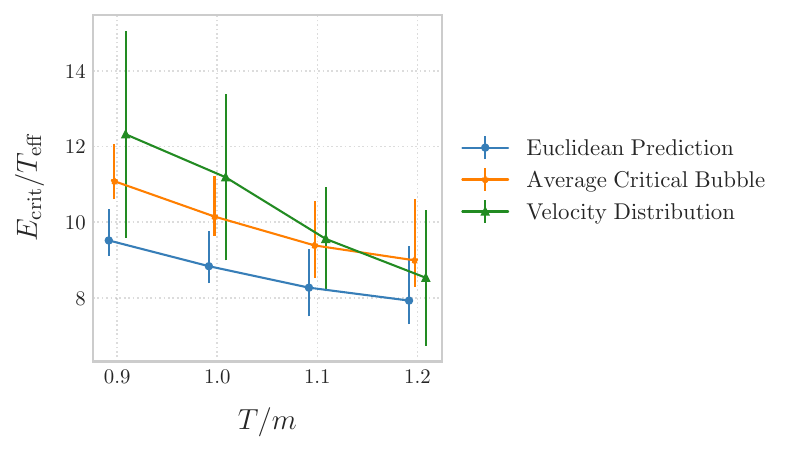}
    \caption{The energy of the critical bubble solution, as measured from three different lattice observables and compared against the Euclidean prediction shown in blue. The prediction is given by Eq.~\eqref{eq:critical_energy_lattice} computed for $\jj_{\crit}$ corresponding to the solution to the static equation of motion Eq.~\eqref{eq:statsphaleron}. The orange line corresponds to the critical field profile obtained empirically from the average bubbles, as explained in Section~\ref{subsec:averagebubble}. The error bars on the blue and orange are proportional to the systematic uncertainty on the effective temperature, as explained in Section~\ref{subsec:powespec_evol}. Finally, the green curve is obtained from taking the inverse of the measured values for $\ev{v_\com^2}$. Here the error bars are the same as in Fig.~\ref{fig:comparison_vcom_pred}, percentage-wise. The data points have been offset slightly along the horizontal axis to help with visualizing the error bars.}
    \label{fig:critenergy_comparison}
\end{figure}

We can compare the observables we introduced up to this point by looking at how they relate to the critical bubble action $B = E_\crit/T_\eff$. In Section~\ref{subsec:decayrate} we have already shown that in the effective IR temperature interpretation we find good agreement with the Euclidean expectation for the decay rate. Here we show that the measurements of the average critical bubble and the velocity distribution are also compatible with the theoretical prediction. The critical bubble energy is
\begin{equation}\label{eq:critical_energy_lattice}
    E_{\crit}(\jj) = \sum_{r}^L \frac{1}{2} \Pi_{\crit}^2 + \frac{1}{2} \left(\partial_{r} \jj_{\crit} \right)^2 + \jj_0^2 V(\jj_{\crit}) - \jj_0^2 V(\jj_{\fv}).
\end{equation}
The kinetic energy term is zero for the static Euclidean solution. Taking the Euclidean solution $\bar{\jj}_{\crit}$ computed for the bare potential (used as initial condition for the left panel of Fig.~\ref{fig:bare_sphaleron_and_subcritical} and plotted in dashed black in Fig.~\ref{fig:critical_profile_comparison}), we obtain a baseline value for the critical energy. Dividing by the empirical effective temperature of each ensemble determined in Section~\ref{subsec:powespec_evol}, we arrive at our theoretical predictions for the action. The predictions are plotted in blue in Fig.~\ref{fig:critenergy_comparison}. Next, we can estimate the energy of the empirical average critical bubble by taking the average field, gradient and momentum profiles on the critical slice shown in Fig.~\ref{fig:critical_profile_comparison} and integrating Eq.~\eqref{eq:critical_energy_lattice}. Repeating the exercise for all four ensembles differing by the value of $T$, we obtain the points shown in orange in Fig.~\ref{fig:critenergy_comparison}. Lastly, according to Eq.~\eqref{eq:varvel_pred}, the variance of the velocity distribution is a direct measurement of the critical bubble action. This is shown in green. Overall, we find good agreement between all measurements in each ensemble.

We have implicitly assumed that the critical energy is independent of temperature, or even effective mass. We postpone a more detailed analysis of the effects related to the running of these variables on the bubble solution to future work. However we note that this approximation is supported by our empirical finding that the average bubble as well as the time evolution of the field and conjugate momentum one-point functions shown in Fig.~\ref{fig:evol_xn_vs_t} are identical, up to statistical error bars, between the four different ensembles. This is in spite of the fact that the averaging is done with respect to different set of parameters $(\bar{\phi}, R)$ (see Section~\ref{subsec:averagebubble}).

\section{Discussion and conclusions}\label{sec:Conclusion}

In this paper we have identified the center-of-mass velocity distribution of bubbles, the ensemble-averaged bubble in the nucleation rest frame, and oscillon precursors as promising observables for vacuum decay. We investigated the properties of these observables using ensembles of classical simulations in 1+1 dimensions. The initial conditions for the simulations are drawn from a non-equilibrium distribution over phase space which we took to be Bose-Einstein (as opposed to the equilibrium Rayleigh-Jeans spectrum). Generalizing our results, we expect bubbles to have a distribution of center-of-mass velocities in any number of spatial dimensions and for any ensemble defining the initial conditions. The distribution likely depends on the ensemble from which the initial conditions are drawn, but it is reasonable to expect that it is given by the Boltzmann factor involving the total bubble energy if the ensemble is close to thermal. This will be a topic of future investigation. We also expect that oscillon precursors play a role in thermal bubble nucleation in any number of spatial dimensions. Here, there could be interesting phenomenology related to the potential, since this determines the properties of oscillons. There may also be dependence on the ensemble of initial conditions, since the oscillons are infrared sensitive objects. For the ensemble-averaged bubble, there may also be important differences that depend on the potential and the ensemble of initial conditions. For example, vacuum decay at zero temperature is not described by the static thermal critical bubble, but by the $O(4)$-invariant Euclidean bounce, which gives rise to an expanding bubble in real time. Can this be the average bubble observed in semiclassical simulations such as those performed in~\cite{braden-newsemiclassical}?

Our work could also have interesting implications for a number of phenomenological scenarios involving first-order phase transitions. Models of electroweak baryogenesis involve a first-order phase transition (see \eg~\cite{Morrissey_2012} for a review). The terminal velocity of the bubble wall through the primordial plasma is a crucial element of these models, determining if a sufficient baryon asymmetry can be accumulated. Incorporating the velocity distribution outlined in this paper could have implications for this calculation if the expected velocity of bubbles is comparable to the terminal wall velocity which is typically non-relativistic and can be as low as $v \sim \mathcal{O}(.1)$~\cite{John_2001, Moore_1995, Moore_1995b} (though in some models can be close to the speed of light, see \eg~\cite{Cline:2020jre}). Our results imply that the root-mean-square bubble center-of-mass velocity is $\sqrt{\ev{v_\com^2}} \sim 1/\sqrt{B}$, where $B \gg 1$ is Euclidean critical bubble action controlling the false vacuum decay rate. We can estimate $B$ by requiring that the phase transition occurs when $\Gamma \sim H(T)^4$, which for $H^2 \sim T^4/M_{\rm pl}^2$ and assuming $\Gamma \sim T^4 B^{3/2} e^{-B}$ (appropriate in 3+1 dimensions) yields $B \sim 150$ for a temperature of order $\rm TeV$. The expected root-mean-square velocity of a bubble during electroweak baryogenesis is therefore $\sqrt{\ev{v_\com^2}} \sim 0.1$ -- comparable to the terminal wall velocity! This simple estimate undoubtedly misses important physical effects, but it certainly motivates the inclusion of bubble velocities in these models. Note that in existing simulations of electroweak baryogenesis \eg~\cite{Hindmarsh_2015}, bubbles are inserted by hand and do not include this effect. These simulations could be augmented to include the velocity distribution outlined in this paper. The velocity distribution may also have implications for the spectrum of stochastic gravitational waves produced during electroweak baryogenesis or other early universe phase transitions observable by LISA~\cite{auclair2022cosmology}.

The oscillon precursor to bubble nucleation could also have implications for baryogenesis. This is because the oscillon core can sample regions of a symmetry-breaking phase. The role of oscillons in electroweak baryogenesis was discussed in~\cite{Riotto_1996}, whose title reflects their conclusion: oscillons are not present during an electroweak phase transition. Here, we have shown that whenever bubbles are present, so are oscillons. This motivates revisiting the question of whether an oscillon precursor could contribute significantly to the dynamics and outcome of baryogenesis.

Early work on oscillons showed that their presence could affect the decay rate of a false vacuum~\cite{Gleiser:2007ts}. Here, we highlight that for thermal decay they are an essential component of the nucleation process. What is their role in the Euclidean formalism -- are they implicitly captured in the saddle point corresponding to the critical bubble? The answer to this question could have implications for the decay rate computation at zero temperature as well: is vacuum decay preceded by a `virtual' oscillon, or is vacuum decay fundamentally different in this respect? We hope to investigate these, and other questions, in future work. 

\acknowledgments 

We thank Thomas Billam, Jonathan Braden, August Geelmuyden, Alex Jenkins, Ian Moss, Hiranya Peiris, Andrew Pontzen, Andrey Shkerin, Vitor Barroso Silveira and Silke Weinfurtner. MCJ is supported by the Natural Sciences and Engineering Research Council of Canada (NSERC) through a Discovery grant. The work of SS is supported by the Natural Sciences and Engineering Research Council (NSERC) of Canada. This research was supported in part by Perimeter Institute for Theoretical Physics. Research at Perimeter Institute is supported by the Government of Canada through the Department of Innovation, Science and Economic Development Canada and by the Province of Ontario through the Ministry of Research, Innovation and Science. Simulations were performed using a version of the \textit{1d-scalar} code~\cite{BradenGit}, and carried out on the Symmetry HPC cluster at the Perimeter Institute.

\appendix

\section{Numerical Convergence Tests}\label{app:A}

In this Appendix, we present the convergence tests conducted to validate our simulations. During nucleation, when the non-linearities are strongest -- particularly near the bubble walls -- some precision loss is expected in the numerical integration scheme. A suitable convergence test involves verifying whether bubbles nucleate at the same space-time coordinates across different simulation parameters. We summarize our findings below. Our fiducial parameter set is shown in Table~\ref{table:1} and represents the values used to obtain the results presented in the main text. In this Appendix, we will refer to these values by the subscript $\cdot_{\rm fid}$. For $100$ simulations at $T/m=0.9$, we varied the integration time step $\dd t$ and lattice spacing $\dd r$ with respect to their fiducial values, and confirmed that the nucleation dynamics were not affected by these changes.

For the first test, we varied the integration time step $\dd t$ over the range $m \dd t = \left[ 10^{-3}, \right. 6.5\times10^{-4}, 5\times 10^{-4}, \left. 4\times10^{-4} \right]$, keeping all other parameters fixed and sampling precisely the same initial conditions in both field and conjugate momentum. The largest value equals twice the fiducial integration time step, $m \dd t_{\rm fid} = 5\times 10^{-4}$. Despite this, all nucleation events occurred at the same coordinate location. 

\begin{figure}[t!]
    \centering
    \includegraphics[width=1\textwidth]{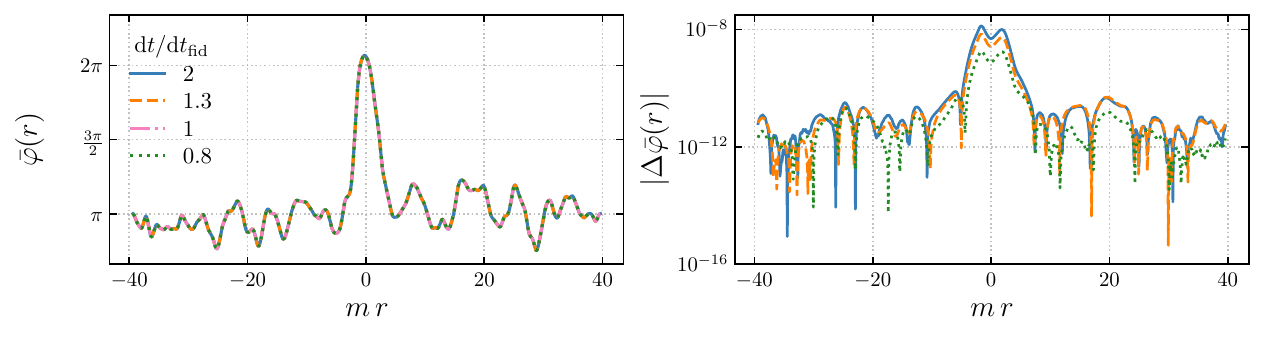}
    \caption{Convergence test results for varying integration time steps $\dd t$. The left panel shows the field profiles, $\bar{\varphi}(r)$, at a fixed time after the bubble's appearance, computed using four different integration time steps. The curves are indistinguishable from the fiducial case, where $m\dd t_{\rm fid} = 0.0005$. The right panel shows the absolute value of the difference between the field profiles and the fiducial realization, $\left| \Delta \bar{\varphi}(r) \right|$. This demonstrates that simulations performed with the fiducial simulation parameters have converged. A similar behavior is observed for the conjugate momentum.}
\label{fig:convergence_test_changing_dt}
\end{figure}

We observed a precision loss of order $10^{-8}$ in the field and momentum, near the bubble walls. Precision here refers to the relative change in field amplitude at fixed coordinates $(r, t)$ between simulations with identical initial conditions but different integration time steps. The left panel of Fig.~\ref{fig:convergence_test_changing_dt} shows the field profiles at the same instant after the bubble's appearance, taken from simulations obtained with four different integration time steps; the curves are indistinguishable. The right panel displays the absolute value of the difference between the field profiles relative to the fiducial realization, denoted by $\left| \Delta \bar{\varphi}(r) \right|$. At its peak, and for the largest time step $\dd t$, this reaches values of order $10^{-8}$ around the nucleation region. This confirms the robustness of the integration scheme.

In the second test, we examined the effects of changing the spatial resolution by doubling and quadrupling the number of lattice sites $N$, while keeping all other parameters fixed, including the box size $L$ and the integration time step $\dd t_{\rm fid}$. The resolution scales were $\dd r/m = 0.08, 0.04$ and $0.02$, respectively. In the fiducial case with $\dd r/m = 0.08$, we set the cutoff for the number of sampled frequencies in the initial field and momentum to the Nyquist limit, $n_{\rm nyq} = N/2 = 512$. Increasing $N$ raises $n_{\rm nyq}$, but we kept $n_{\rm cut} = 512$ constant to verify that the modes not sampled (those between $n_{\rm cut}$ and $n_{\rm nyq}$) did not significantly impact the realizations. In this test, the bubbles also nucleated at the same space-time location. The observed precision loss was approximately $1\%$, due to power in the initial conditions cascading to initially unoccupied modes, which induced UV fluctuations that locally altered the field amplitude. Nonetheless, as previously stated, the high-frequency, out-of-equilibrium UV bath did not significantly affect the IR-relevant dynamics of bubble formation. Note that to measure the amplitude change between simulations with a different number of lattice sites, we must first interpolate the field on the finer grid. We found that the amplitude difference was similar in magnitude regardless of the choice of interpolation scheme.

\begin{figure}[t!]
    \centering
    \includegraphics[width=0.55\textwidth]{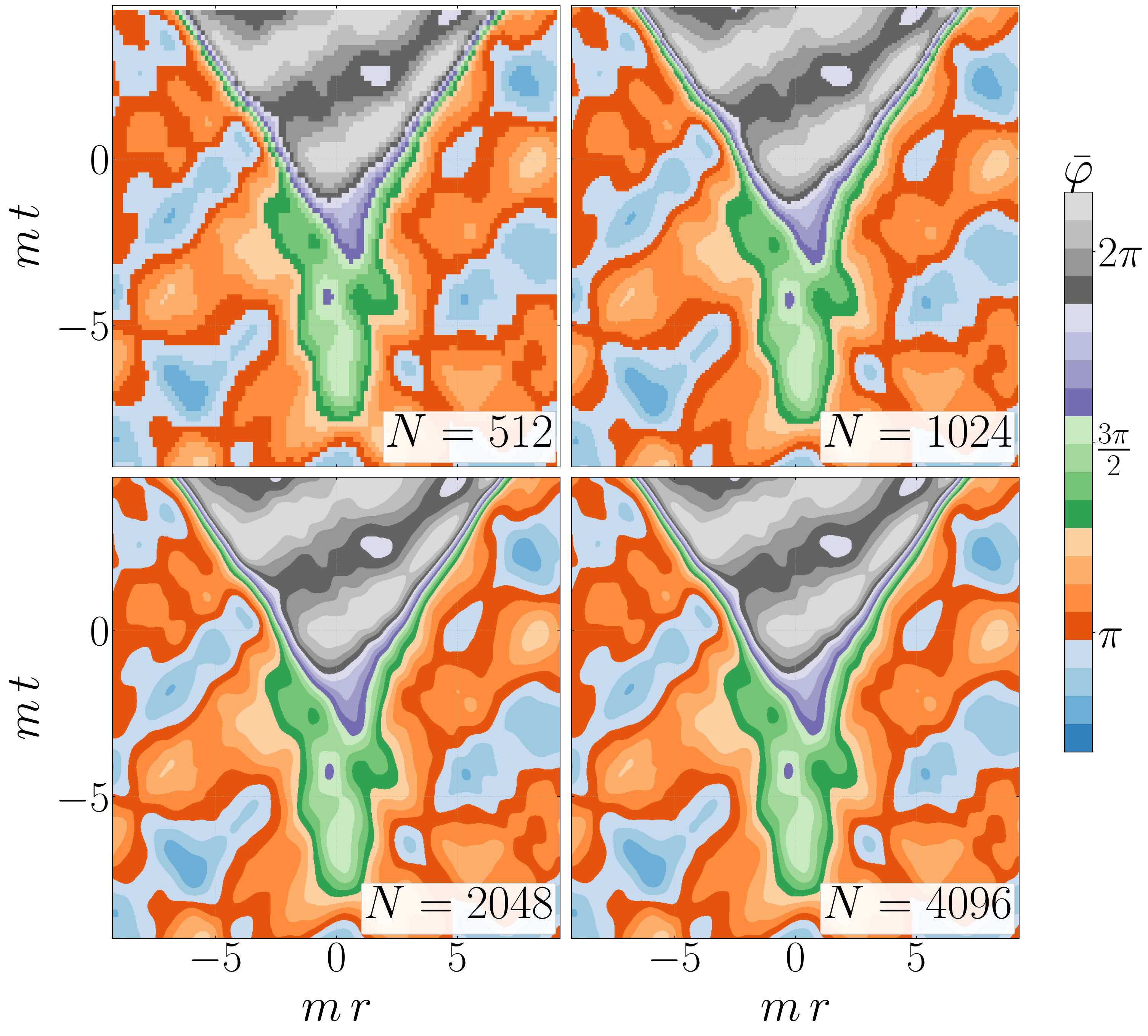}
    \caption{Convergence of bubble nucleation dynamics across different lattice resolutions. The panels show simulations with varying numbers of lattice sites $N$, corresponding to different lattice resolutions. For the three cases with largest $N$, we initialized $512$ modes with the same initial amplitudes at $t=0$; for the $N=512$ case, only the first $256$ Fourier modes were initialized. The color map represents the field amplitude $\bar{\varphi}$ normalized across all four panels. Note that the image pixel size changes with the lattice resolution. All bubbles nucleate at the same space-time location, indicating consistent field dynamics across different spatial resolutions, with only minor amplitude fluctuation differences. This demonstrates that the dynamics has converged for our fiducial set of parameters.}
\label{fig:convergence}
\end{figure}

The final test involved taking the initial conditions for our fiducial set of simulations (namely, the Fourier amplitudes of each mode in both field and conjugate momentum), and re-initializing a new set of simulations on a lattice with $N = N_{\rm fid}/ 2 = 512$ sites and $n_{\rm nyq} = 256$, discarding half of the modes (those with higher frequencies). The new simulations had a lattice resolution scale $\dd r = 2 \dd r_{\rm fid} = 0.16/m$ and a UV scale of $k_{\rm UV} = \dd k N/2 = 20.5 m$. Here as well, across all simulations probed, the nucleation coordinates were unchanged compared to the fiducial case.

An illustration of a simulation initialized on four different grids, as described above, is shown in Fig.~\ref{fig:convergence}. The fiducial case corresponds to the top-right panel, where $N=1024$. In all four cases, the bubble nucleates at $m t = 325$ from the start of the simulation and at the same position $m r$. Not only does the bubble appear at the same spot, but the entire field dynamics is similar in all cases. We conclude that our physical and lattice parameters capture all relevant dynamics for bubble nucleation.

\section{Velocity measurement and boosting to the rest frame}\label{app:BB}

\begin{figure}[t!]
    \centering
    \includegraphics[width=0.7\textwidth]{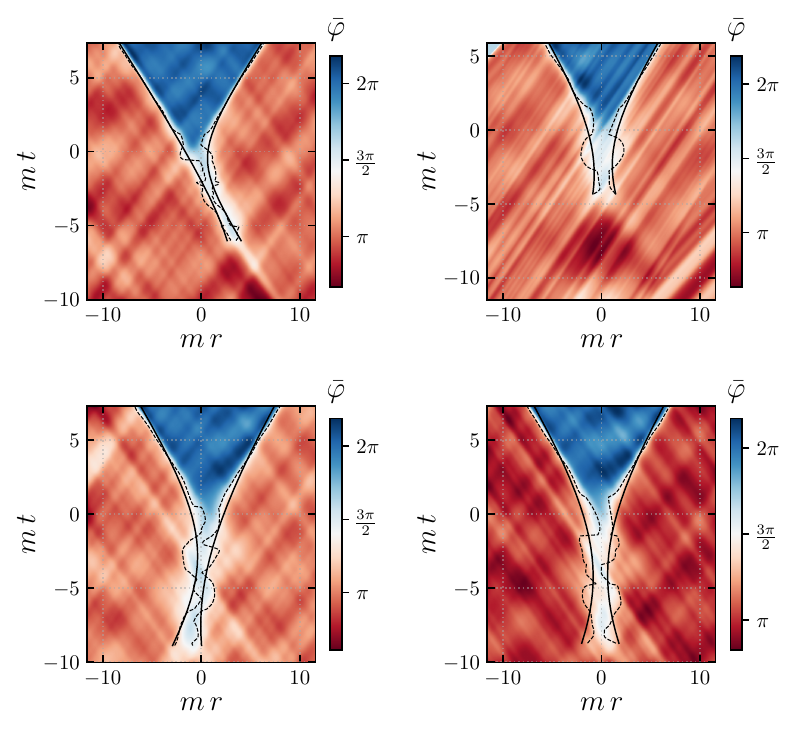}
    \caption{The spacetime diagrams on the left-hand side depict two bubbles as detected in their original realizations, while on the right we show the same bubbles at rest in the simulation frame, after going through the de-boosting procedure. The top bubble was detected to have the center-of-mass velocity $v_\com= -0.78$, while the one in the bottom panels was measured to have $v_\com= 0.26$. The greater the center-of-mass velocity is in absolute value, the more Lorentz contracted the bubble appears in the original simulation. The dashed line represents the wall trajectory found by fitting the field at each time slice to the wall profile in Eq.~\eqref{eq:bub_profile}. The solid black line is the hyperbolic best-fit to the dashed line curve, from which the instantaneous wall velocities are obtained. Stronger boosts lead to stronger field distortion and the introduction of a cutoff for the late-time expansion (\eg from an original square grid to an area-conserved boosted diamond). This can be seen in the top left corner of the diagram in the top right panel. In the same image a large amplitude coherent fluctuation is seen as the bubble precursor. The bubble precursor as well as the neck of the bubble in the initial stage of expansion are noticeably thinner in the left panel than on the right.}
\label{fig:examples_deboost}
\end{figure}

In this appendix, we describe in detail the procedure for measuring the center-of-mass velocity of bubbles identified in simulations. The first step is to map the trajectories of the expanding bubble walls, denoted $r_{\rm L/R}(r,t)$. To achieve this, we model the field amplitude at each time step using the expression 
\begin{equation}\label{eq:bub_profile}
    \bar{\jj}(r,t={\rm const}) = \left( \tanh{ \frac{r-r_{\rm L}}{w_{\rm L}}} + \tanh{\frac{r_{\rm R}-r}{w_{\rm R}}}\right) \frac{\bar{\jj}_{\fv}}{2} + \bar{\jj}_{\fv},
\end{equation}
where $r$ is a coordinate that spans the lattice and $w_{\rm L,R}$ represent the thickness of each wall. In this expression, $r_{\rm R,L}$ are the best-fit coordinates of the wall centers. The hyperbolic tangent profile provides an excellent fit for the shape of the domain walls. Starting at $t\gg t_{\crit}$, we trace the evolution backward in time towards the `fuzzy' nucleation region, finding the best-fit values for $r_{\rm R,L}$ at each step. Two examples of such trajectories are illustrated by the dashed curves in Fig.~\ref{fig:examples_deboost}.

To mitigate the effect of fluctuations, we select several initial guesses for the start values of the parameters in Eq.~\eqref{eq:bub_profile} corresponding to wall amplitudes in the range $\bar{\jj}_\fv + 1.5\sigma_{\bar{\jj}}$ and $\bar{\jj}_\fv + 3\sigma_{\bar{\jj}}$. These start values in general will generate different wall trajectories. At each step $t$ beyond the starting slice, the initial guess for the best-fit parameters is updated to the values obtained at step $t+\dd t_{\rm out}$.

With the independent trajectories $r_{\rm L,R}$ obtained in this manner, each wall is fitted to a hyperbolic function $r_{\rm L,R}(t) = \pm \sqrt{a_1 + (t - a_2)^2} + a_3$ with free parameters $a_1, a_2, a_3 \in \mathbb{R}$. We obtain a bundle of hyperbolas expanding at roughly similar rates. From here, the instantaneous wall velocity is simply the tangent curve $v_{\rm L,R}(t) = \partial_{t} \, r_{\rm L,R}$. By fitting first to a hyperbolic trajectory, we ensure that the $|v_{\rm L,R}| \leq 1$ at all times, as well as smooth out the effect of fluctuations. This can be seen by comparing the dashed and solid lines in Fig.~\ref{fig:examples_deboost}.

In the rest frame, the velocities of the left- and right-moving walls, $v_{\rm wall}(t)$, are equal. In a boosted frame, $v_{\rm L,R}(t)$ are related by a gamma factor that is a function of $v_\com(t)$. Since at every time-step we have two equations with two unknowns, the instantaneous center-of-mass velocity is fully determined. Its value \emph{at nucleation} is chosen as the instantaneous $v_\com(t)$ that minimizes the residual $|v_\com(t) - v_{\rm wall}(t)|$. Once again, this is because at nucleation the expectation is that the walls start off at rest.

For each wall trajectory determined by different initial values of the fit parameters in Eq.~\eqref{eq:bub_profile}, we obtain a different measurement for the $v_\com(t)$ that minimizes the residual. Some trajectories may fail to provide a numerical estimate for $v_\com(t)$; for example, in many failed cases, no hyperbolic fit is found. In general, since the wall amplitude spans a small range in $\bar{\jj}$ of only $1.5\sigma_{\bar{\jj}}$, the values of the center-of-mass velocities at nucleation obtained from the different hyperbolas will differ by less than $10\%$. We use the average of all these values as the final result of the measurement.

We may refer to the center-of-mass velocity measured in this way as the \emph{deterministic velocity}. In the absence of fluctuations, this procedure could be applied once, and the true center-of-mass velocity would equal the deterministic velocity. However, the presence of fluctuations introduces uncertainties in the measurement. In the worst case, it can lead the algorithm down a wrong path, and boost the bubble into a \emph{more} relativistic frame. Therefore, we need to iterate over this procedure several times, verifying at each step that we are on the right track.

We test the value obtained for the deterministic velocity by applying a Lorentz transformation with the corresponding boost factor $\gamma(v_\com)$ and measuring the center-of-mass velocity once again. If the newly detected velocity is less than the original, the value is accepted. This procedure is repeated until a residual $\leq 0.03c$ is reached, which serves as a lower bound for the error in the measurement. However, if at any point, the transformation results in a residual that is larger than the applied velocity, that value is discarded, and random velocities are applied until a frame where the bubble is closer to rest is found. These random values are chosen in the interval $0.05 < |v| < 0.2$. The final overall $v_\com$ is the result of relativistic addition of all boost factors that have been accepted, whether random or deterministic.

To affix a de-boosted bubble onto a transformed grid, we first linearly interpolate the field (in the frame where the velocity was most recently measured), and then evaluate it onto the new grid (with coordinates determined via the Lorentz boost). Recall that boosts preserve the spacetime interval but weight the time and space components differently, resulting in a distorted bubble. Note that repeated linear interpolation at each intermediate step introduces noise in each realization. To minimize this noise, once the final $v_\com$ is measured, we apply a single Lorentz boost with this value to the original realization and check again that the output bubble is measured at rest. If the residual satisfies $\leq 0.03c$, the procedure has been completed successfully.

An illustration of this procedure is shown in Fig.~\ref{fig:examples_deboost} for two examples of bubbles at $T/m = 0.9$. The left panel shows the spacetime diagram of the bubbles as they appear in a particular simulation. The right panel displays the final result after applying a Lorentz boost with velocity $v_\com = -0.78$ (top) and $v_\com = 0.26$ (bottom), centered on the nucleation event. The dashed lines represent the measured wall trajectories, while the solid lines correspond to the associated hyperbolic fits in each frame. Notice that, after de-boosting, the left and right wall trajectories are now (almost) symmetric about $r=0$. Additional examples of bubbles before and after the de-boosting procedure are depicted in Fig.~\ref{fig:many_examples_deboost_before_after} in the main text, for the case where $T/m=0.9$.

\bibliography{references.bib}
\bibliographystyle{JHEP}

\end{document}